\documentclass[%
 reprint,
 amsmath,amssymb,
 aps,
]{revtex4-2}

\usepackage[utf8]{inputenc}

\usepackage{todonotes}
\usepackage{amssymb,amsmath,amsbsy,mathtools}
\usepackage{braket}
\usepackage{graphicx,color}
\usepackage{hyperref}
\usepackage{subcaption}
\newcommand{\mH}{\mathcal{H}}

\newcommand{\mO}{\mathcal{O}}

\newcommand{\mR}{\mathcal{R}}

\newcommand{\mT}{\mathcal{T}}
\newcommand{\lb}{\left(}
\newcommand{\ep}{\epsilon}
\newcommand{\rb}{\right)}

\newcommand{\mI}{\mathbb{I}}

\newcommand{\p}{\partial}

\newcommand{\nn}{\nonumber}
\newcommand{\ee}{\end{equation}}

\usepackage{graphicx}

\usepackage{graphicx}% Include figure files
\usepackage{dcolumn}% Align table columns on decimal point
\usepackage{bm}% bold math
\begin{document}

\preprint{APS/123-QED}

\title{Renormalization group and approximate error correction}% Force line breaks with \\
\author{Keiichiro Furuya$^{1}$, Nima Lashkari$^{1,2}$, Mudassir Moosa$^{1}$}
\affiliation{
 ${}^1$ Department of Physics and Astronomy, Purdue University, West Lafayette, IN 47907, USA}
% \author{Nima Lashkari}%
%  \email{lashkari@purdue.edu}
%  \altaffiliation[Also at ]{School of Natural Sciences, Institute for Advanced Study, Princeton, New Jersey 08540, USA}
 %


% \author{Mudassir Moosa}
%  \homepage{http://www.Second.institution.edu/~Charlie.Author}
\affiliation{
  ${}^2$ School of Natural Sciences, Institute for Advanced Study, Princeton, New Jersey 08540, USA}%
% \affiliation{
%  Third institution, the second for Charlie Author
% }%
% \author{Mudassir Moosa}
% \affiliation{%
%  Authors' institution and/or address\\
%  This line break forced with \textbackslash\textbackslash
% }%

% \collaboration{CLEO Collaboration}%\noaffiliation

%              %  but any date may be explicitly specified

\begin{abstract}
In renormalization group (RG) flow, the low energy states form a code subspace that is approximately protected against the local short-distance errors. We motivate this connection with an example of spin-blocking RG in classical spin models. We consider the continuous multi-scale renormalization ansatz (cMERA) for massive free fields as a concrete example of real-space RG in quantum field theory (QFT) and show that the low-energy coherent states are approximately protected from the errors caused by the high-energy localized coherent operators. In holographic RG flows, we study the phase transition in the entanglement wedge of a single region and argue that one needs to define the price and the distance of the code with respect to the reconstructable wedge.
% The trade-off bounds set a bound on the amount of quantum information at a scale. We comment on the connections to the quantum error correction codes in holography and the high-energy states of chaotic quantum systems.
\end{abstract}

\maketitle

%\tableofcontents

\section{Introduction}

Renormalization group (RG) flow is a pillar of the twentieth century physics that has allowed us to study the universal dynamics of the emergent long-range effective degrees of freedom. Quantum error correction (QEC) teaches us how to encode quantum information non-locally to protect against local noise and decoherence. They both involve the physics of states with long-range correlations, and the fixed points of RG flow are intimately tied to error correction codes. We can view the RG as an isometric embedding $W$ of the infrared (IR) states into the ultra-violet (UV) states. In the analogy to error correction codes, this isometry is an encoding map, the IR states are the logical states, and the UV states are the physical states: $W:\mH_{IR}\to \mH_{UV}$\footnote{In general, it suffices to take $W$ to be an approximate isometry. In relativistic theories $W$ can be unitary.}. 
% \KF{In general, it suffices to take $W$ to be an approximate isometry. In relativistic theories $W$ can be unitary.}
The decoding map is $W^\dagger$ and the projection $P=WW^\dagger$ projects the physical Hilbert space down to the code subspace. Irrelevant local perturbations are the noise that the encoding protects against: $W \mathcal{O}_{irrel}W^\dagger\sim P$. In this work, we explore this connection in three examples: 1) the RG flow of classical Ising model, 2) the real-space RG flow of quantum massive free fields realized as continuous Multi-scale Renormalization Ansatz (cMERA), 3) the holographic RG flows.

\begin{figure}
    \centering
    \includegraphics[width=0.4\linewidth]{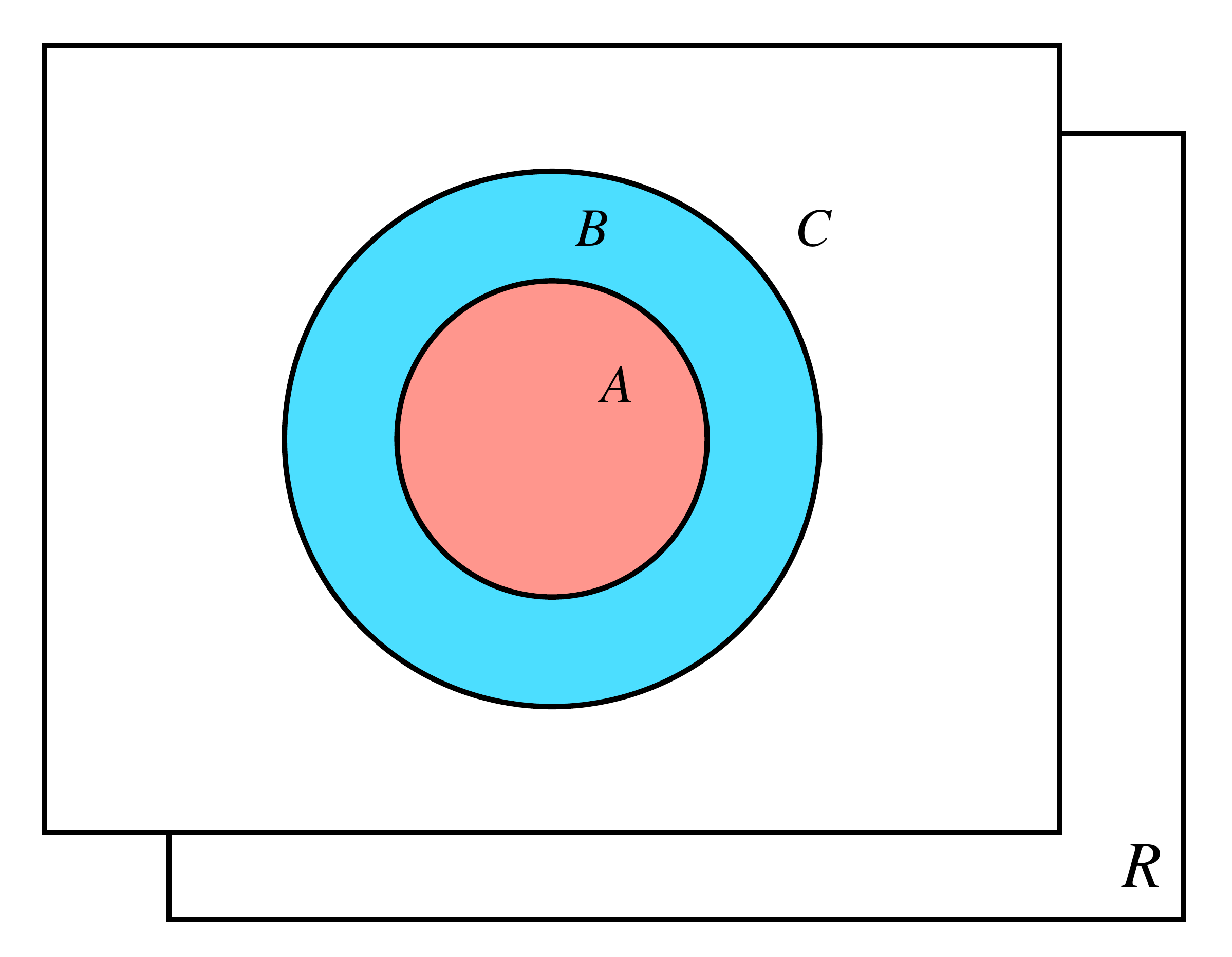}
    \caption{\small{The spatial region is partitioned by $ABC$ with purifying system $R$. The local erasure $tr_A$ acts on the red-colored region A. The blue-colored region is the spatial domain of a local recovery ma, $\mR_{B}^{AB}$.}}
    \label{fig1}
\end{figure}

 %\footnote{For the readers less familiar with relativistic quantum field theory (QFT) we mention the following examples of each type: Quantum Chromodynamics (QCD) is of type one, because under the RG the coupling grows and we end up with massive glueballs in the infrared. Quantum Electrodynamics (QED) and models with spontaneous breaking of a continuous symmetry are of second type. Photons and the Nambu-Goldstone boson are the surviving massless mode in the IR. Non-Abelian Yang-Mills theory coupled to fermions in the conformal window flows to the Bank-Zach fixed point in the IR, and the $\lambda \phi^4$ scalar theory in $D=4-\epsilon$ dimensions flows to the Wilson-Fisher fixed points. These are both examples of type three.
%}. 
We obtain an exact error correction code at the IR end point of an RG flow if we have degenerate ground states which can occur either due to spontaneous symmetry breaking or topological order. Spontaneous breaking of discrete or continuous symmetries leads to classical error correction codes. %\footnote{For instance, the two-dimensional Ising model at low temperature breaks the $\mathbb{Z}_2$ symmetry spontaneously by forming long-range ordered ferromagnetic states $\ket{00\cdots 0}$ and $\ket{1\cdots 1}$. This is a classical repetition code that corrects for local bit flips ($\sigma_X$ Pauli matrix). However, the existence of a local order parameter that distinguishes different code states precisely implies that the local density matrices are distinguishable and one cannot correct local quantum errors ($\sigma_z$ Pauli matrix).}. 
% Spontaneous breaking of discrete and continuous symmetries leads to classical codes in the subspace of vacua.\footnote{Classical codes can be understood as the special case of QEC where the correctable algebra is Abelian.}
We obtain QEC codes in the vacuum subspace when we have topological order. A quantum system with degenerate vacua is said to have topological quantum order if the reduced density matrix of any ball-shaped subregion $A$ is the same in all the ground states.
% \footnote{If we call the subspace spanned by the ground states the code subspace and purify it in a reference system $R$, the above definition is equivalent to the decoupling condition $I_\rho(A:R)=0$, which is the definition of an exact erasure-correction quantum code.}.\KF{If we call the subspace spanned by the ground states the code subspace and purify it in a reference system $R$, the above definition is equivalent to the decoupling condition $I_\rho(A:R)=0$, which is the definition of an exact erasure-correction quantum code.}
We will see that in the absence of topological order, for the low energy states the errors on of small enough subregions $A$ are approximately protected.
Similar to \cite{flammia2017limits}, we find that the QEC is local. This means that that the recovery map $\mathcal{R}$ that corrects the errors on $A$ is localized on $AB$ where $B$ is some small region that surrounds $A$; see figure \ref{fig1}.\footnote{One way to obtain a local QEC is to consider a code where we can correct simultaneous errors on $A$ and $C$; see figure \ref{fig1}.}%\footnote{On a lattice, LQECC are represented by the following four parameters $[[n,k,d]]_q$, where at each site of the lattice we have a $q$-level system, $n$ is the number of physical sites,  $k$ is number of the logical ``$q$-bits'', and $d$ is the distance of the code. The distance is defined to be size of the support of the smallest logical operation. In QEC, there is a trade-off between the rate of logical q-dits we encode $k/n$ and the distance $d$. For a fixed rate the distance has an upper bound. For instance, for local commuting codes in $D$-spatial dimensions we have the trade-off bound $k d^{2/(D-1)}\leq c n$ where $c$ is a constant \cite{bravyi2010tradeoffs,haah2012logical}. A subclass of LQEC codes constructed out of local commuting projections includes the topological ordered systems such as Kitaev's quantum double model and the Levin-Wen string-nets \cite{kitaev2003fault,levin2005string,cui2020kitaev}. In continuum topological quantum field theories, it is expected that the states prepared using Euclidean path-integral lead to QEC codes \cite{salton2017entanglement}. In the case of Abelian Chern-Simons theory, the ground subspace is a stabilizer code. Since topological order is stable, small perturbations of commuting projection codes leads to approximate error correction codes \cite{flammia2017limits}.}.

%  \KF{The spatial region is partitioned by $ABC$ with purifying system $R$. The local erasure $tr_A$ acts on the red-colored region A. The blue-colored region is the spatial domain of a local recovery ma, $\mR_{B}^{AB}$.}

The RG flows that end with scale-invariant states are also intimately tied to approximate error correction codes. This connection holds both for classical and quantum RG flows. In section \ref{sec:classicalspinblock}, we describe how Kadanoff's block-spin renormalization of a classical system leads to approximate classical error correction codes. 
The quantum analog of spin blocking approach to RG is the Multi-scale Entanglement Renormalization Ansatz.
MERA is a particular class of isometric tensor networks that is well-suited for describing the real-space RG flow \cite{vidal2007entanglement}\footnote{On a lattice, tensor networks provide a convenient graphic notation to study QEC codes \cite{ferris2014tensor}.}.
% \KF{On a lattice, tensor networks provide a convenient graphic notation to study QEC codes \cite{ferris2014tensor}.}. 
MERA describes topological codes \cite{aguado2008entanglement,konig2009exact}, as well as the low energy states of scale-invariant theories. It was argued in \cite{kim2017entanglement} that MERA viewed as an isometric embedding of the logical Hilbert space of states at some IR energy scale in the physical Hilbert space at the UV scale is an approximate QEC code. The encoding of the IR information is approximately protected against the erasure of small local regions in the UV lattice \footnote{The authors of \cite{kim2017entanglement} showed that MERA as an approximate QEC code satisfies the trade-off bound $k d^\alpha\leq cn$ where $\alpha$ is a constant fixed in terms of the size ratio $|AB|/|A|$. See the supplementary material.}. 
% \KF{The authors of \cite{kim2017entanglement} showed that MERA as an approximate QEC code satisfies the trade-off bound $k d^\alpha\leq cn$ where $\alpha$ is a constant fixed in terms of the size ratio $|AB|/|A|$. See the supplementary material.}. 
Here, we argue that this connection generalizes to the RG flow of continuum Poincare-invariant QFTs. As a step in making this connection rigorous, in \cite{furuya2020real}, the operator algebra exact QEC was generalized to arbitrary von Neumann algebras, including the local algebra of QFT. 

In section \ref{sec:cMERA}, we consider continuous MERA (cMERA) for continuum QFT of massive free fields in $2$-dimensions. We use the field coherent states to  encode quantum information locally. We study the RG flow of these code states and show that the low energy coherent states form a local quantum error correction code that is approximately protected against the action of the UV coherent operators in $A$ and $C$; see figure \ref{fig1}.

In section \ref{sec:holography}, we consider a three-dimensional geometry corresponding to a holographic RG flow from a UV to an IR CFT. The entanglement wedge corresponding to a finite boundary interval shows a phase transition which naively suggests that there are finite volume regions in the bulk where no information can be encoded. We resolve this paradox by observing that near the phase transition point the reconstruction wedge is smaller than the entanglement wedge. One needs to define the notions of code price and distance with respect to the reconstructable wedge.
%This implies a modification of the definitions of price and distance in holography. 
% We discuss the implications of the error correction perspective for the algebra of observables of an effective QFT at mass scale $\mu$. In section \ref{sec:O(N)}, we study the RG flow of the excited states of free $O(N)$ model and its connection to the RG flow. In section\ref{sec:null}, . In section \ref{sec:holography}, we give holographic examples of RG flows and QEC codes. 
We conclude with a summary and discussion in section \ref{sec:conclude}.

% If the gapped system has a unique ground state we will not obtain an exact error correction code. However, we still have an approximate error correction code where the operators at some scale above the mass gap are logical and the local operators in the UV are noise. We study examples of such classical and quantum codes in 
% section \ref{sec:EFT}. A mass gap with degenerate ground states can occur either due to spontaneous symmetry breaking or topological order. In section \ref{sec:SSB}, we give examples of classical error correction codes that occur in the IR due to spontaneous symmetry breaking.\NL{Continuous SSB and encoding a harmonic oscillator?}  

\section{Classical spin-blocking}\label{sec:classicalspinblock}

\begin{figure}
    \centering
    \includegraphics[width=0.6\linewidth]{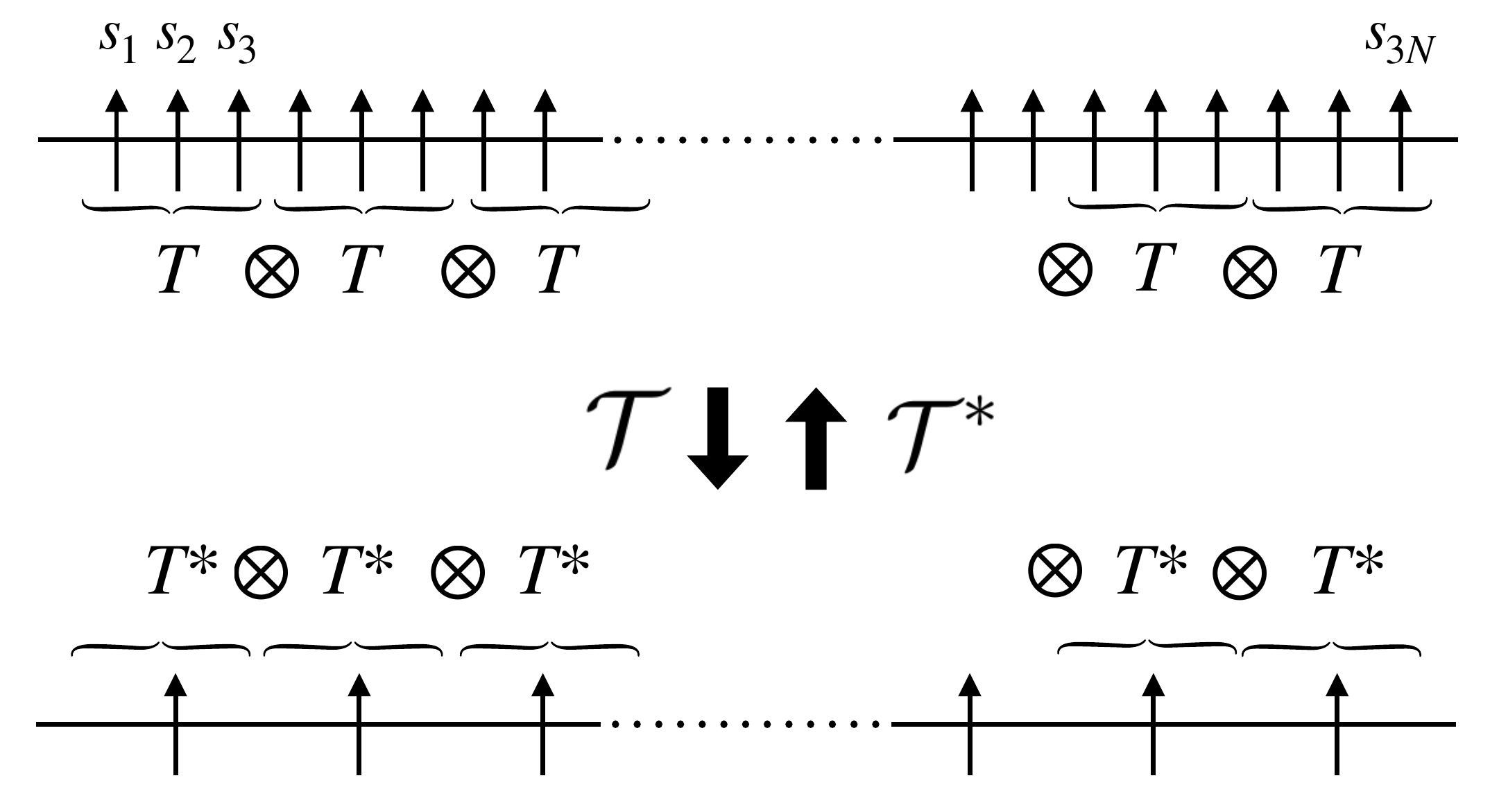}
    \caption{\small{This is a single step coarse-graining on a classical Ising model in one dimension with $3N$ sites. Since $T$ coarse-grains three spins into a single spin, $\mT = T^{\otimes N} $ coarse-grains $3N$ sites into $N$ sites.}}
   \label{fig2}
\end{figure}

In classical physics, the standard real-space RG scheme of a spin systems involves splitting the lattice into blocks and coarse-graining the degrees of freedom localized inside each block. 
The coarse-graining replaces the collection of spins inside each block $\{s_i\}$ with a collective degree of freedom $s'$ \cite{kadanoff1967static,wilson1975renormalization}. Let $T$ be the coarse-graining map. It induces the encoding isometry mapping from the $s'$ variable as the logical state to $\{s_i\}$ variables as the physical states. For example, consider the translation-invariant one dimensional classical Ising model  with $3N$ sites satisfying the periodic boundary conditions; see figure \ref{fig2}. %, a local Hamiltonian
%\begin{eqnarray}
%H(\{s\})=-J\sum_{i=1}^{3N-1} s_i s_{i+1}+h\sum_{i}^{3N-1} s_{i}
%\end{eqnarray}
%and periodic boundary conditions; see figure \ref{fig2}.
%where $s_i$ is the $i^{th}$ spin variable and $i=1,\cdots, 3N$.
%The configuration space is the set of all sequences  $\{s\}=\{s_1s_2\cdots s_{3N}\}$ and the thermal state is the probability distribution
%\begin{eqnarray}
%p(\{s\})=e^{-\beta H(\{s\})}/Z
%\end{eqnarray}
%on this configuration space. 
A simple coarse-graining scheme is the majority vote scheme
\begin{eqnarray}\label{majvote}
s'=\begin{cases}
    +1       & \quad \text{if} \: s_1+s_2+s_3\geq 0\\
    -1  & \quad \text{otherwise}\ .
    \end{cases}
\end{eqnarray}
%It is convenient to represent states as vectors $\ket{s_1s_2s_3}$ and think of the coarse-graining map as a matrix $T$ describing a classical information channel acting on probability vectors \footnote{The matrix elements are $\braket{s'|T|s_1s_2s_3}$. In our example, the map $T$ is $T=\big(\begin{smallmatrix}
%  1 &1&1&0&1&0&0&0\\
%  0 &0&0&1&0&1&1&1 
%\end{smallmatrix}\big)$.}. 
In this model, the encoding isometry is the normalized transpose map $T^*\equiv T^T/4$. %Denoting a physical state of each block as $\ket{s_1s_2s_3}$, the code subspace is spanned by the two states $\ket{T^*(\pm 1)}$ \footnote{The encoded states are
%$\ket{T^*(\pm 1)}=\frac{1}{4}(\ket{\pm1,\pm1,\pm1}+\ket{\pm1,\pm1,\mp1}+\ket{\pm1,\mp1,\pm1}+\ket{\mp1,\pm1,\pm1})$.}.
A simple model of local noise is a one-site bit flip error $\pm 1\to \mp 1$ with probability $p$. Its action on the $i$th bit is given by the symmetric binary channel $G_i=\big(\begin{smallmatrix}
  1-p & p\\p&1-p
\end{smallmatrix}\big)$. %In the physical space this error is represented by $G_i\otimes \mI_{\backslash i}$. 
The noise propagates to the logical bits $s'$ and acts as the matrix
\begin{eqnarray}\label{logicalerrror}
\Phi(G_i)= \mathcal{T}(G_i\otimes \mI_{\backslash i})\mathcal{T}^* =\begin{pmatrix}
1-\frac{p}{2}&\frac{p}{2}\\
\frac{p}{2} & 1-\frac{p}{2}
\end{pmatrix}\ 
\end{eqnarray}
where $(G_i\otimes \mI_{\backslash i})$ is the local error in the UV, and $\mT = T^{\otimes N}$. By translation invariance of the code states, the other local errors $G_j (j\neq i)$ also lead to the same logical error matrix above. The key observation is that the error after coarse-graining (\ref{logicalerrror}) is the same as the original error but weaker, because the probability of bit-flip is cut in half. In this case, the noise is an eigen-operator of the coarse-graining map $\Phi(\cdot)=\mT(\cdot )\mT^*$ with eigenvalue $1/2$.

In real-space RG, after we repeat the coarse-graining map $n$ (large number) times to flow from the short-distances to long-distances, %In the case of the above model with $3N$ sites, if we start with the local error map $G_i$ on some site $s_{i}$, after the first step the error is
%\begin{eqnarray}
%\Phi(G_i)= \mathcal{T}(G_i\otimes \mI_{\backslash i})\mathcal{T}^*
%\end{eqnarray}
%where $(G_i\otimes \mI_{\backslash i})$ is the local error in the UV, and $\mT = T^{\otimes N}$.
%After $n$ steps of coarse-graining 
the errors are exponentially weaker
\begin{eqnarray}
\Phi^n(G_i)=\begin{pmatrix}
1-2^{-n}p&2^{-n}p\\
2^{-n}p & 1-2^{-n}p
\end{pmatrix}\ .
\end{eqnarray}
As we flow from the very short distances (UV) to very long distances (IR) the local errors are expected to decay exponentially fast $\lim_{n\to \infty}\Phi^n(G_i)=\mI$.

Next, consider the non-local error $G_1\otimes G_2\cdots \cdot \otimes G_k$ that corrupts $k$ adjacent sites. After one level of coarse-graining it corrupts $\lfloor (k-2)/3\rfloor+2$ sites. After each step of coarse-graining the support of non-local errors shrinks almost by a factor of three, until it becomes local at which point the above analysis applies. This logic extends to arbitrary $k$-site error model. There are two stages to the renormalization of any error of finite support in the UV Hilbert space. In the first stage, the support of the operator shrinks monotonically. In the second stage, the error becomes exponentially weaker \cite{kim2017entanglement}. Deep in the IR the RG flow is an approximate classical error correction code in the trivial sense that $k$-local errors are highly unlikely to corrupt the encoded data.

% \KF{This is a single step coarse-graining on a classical Ising model in one dimension with $3N$ sites. Since $T$ coarse-grains three spins into a single spin, $\mT = T^{\otimes N} $ coarse-grains $3N$ sites into $N$ sites.}

The RG flow map $\Phi(\cdot)=\mathcal{T}(\cdot)\mathcal{T}^*$ is a classical channel and all its eigenvalues have norm that is less than or equal to one \footnote{The asymmetric binary channel can be expanded as $G_1=p_1\big(\begin{smallmatrix}
  1-p_1 & p_1\\p_2&1-p_2
\end{smallmatrix}\big)=\mI+p_1 G_++p_2 G_-$ where both $G_+$ and $G_-$ are eigenoperators of the RG map with eigenvalue $1/2$ and the identity map is invariant.}.
% \KF{The asymmetric binary channel can be expanded as $G_1=p_1\big(\begin{smallmatrix}
%   1-p_1 & p_1\\p_2&1-p_2
% \end{smallmatrix}\big)=\mI+p_1 G_++p_2 G_-$ where both $G_+$ and $G_-$ are eigenoperators of the RG map with eigenvalue $1/2$ and the identity map is invariant.}
In our example, there are no one-site errors that are left invariant under the RG map. The only fixed point of the RG map corresponds to acting with the noise at every single site. The support of such an operator never shrinks to one-site. %\footnote{The ground states of the one-dimensional nearest neighbor Ising model form an exact repetition code. A simultaneous $\mathbb{Z}_2$ flip on all sites is a logical operation which takes us from one code state to another in the repetition code.}.
After a large number of RG steps, the encoding is $(\mathcal{T}^*)^n$ and the largest eigenvalue of the RG map controls how well this approximate error correction code protects classical information.

\section{continuous MERA}\label{sec:cMERA}

The intuitive discussion above generalizes to the renormalization group flow of quantum systems with local Hamiltonians. In a gapped system, the RG flow becomes trivial at scales above the correlation length. Since we are interested in repeating the RG map many times, we focus on the real-space RG in critical systems. For a lattice theory, in the IR, the RG map can be viewed as an encoding isometry $W:\mH_{IR}\to \mH_{UV}$. Consider MERA that goes from deep in the UV to the most IR layer with $n$ local sites. We label the layer deep in the IR with $s=0$, each step of coarse-graining $s$ goes up by one, and deep in the UV $s$ is a large negative number and we have $2^{-s}n$ sites. In the thermodynamic limit $n\to \infty$ the MERA network can go on forever and the range of $s$ becomes $(-\infty,0)$. The isometry $W$ is comprised of two layers, first a layer of local isometries $V\otimes \cdots \otimes V$, and second a layer of local unitaries $U\otimes \cdots \otimes U$ called the disentanglers. The layer of local isometries is the quantum analog of $\mathcal{T}^*$ map above. In real-space RG, the disentanglers correctly remove the entanglement of a ground state wave-function $\ket{\Omega^{(s)}}$ at each scale. The state deep in the IR has zero correlation length and no spatial entanglement $\otimes_x \ket{\Omega(x)}$. It was shown in \cite{kim2017entanglement} that MERA viewed as an encoding of the IR information in the UV state is an approximate local quantum error correction code.

A generalization of MERA to continuum theories (cMERA) was proposed in \cite{haegeman2013entanglement}. Similar to the discrete case, cMERA is an isometric map that takes the states of a theory with zero correlation length deep in the IR and prepares the low energy states of a QFT (or CFT) in the UV. The IR ground state is taken to be a state $\ket{\Omega^{(0)}}$ with no real-space entanglement.

Here, we follow \cite{zou2019magic} to construct the cMERA state $\ket{\Omega^{(s)}}$ at scale $e^{-s} \Lambda$ with a choice of scale independent entangler $K$ and the non-relativistic scaling transformation for a free massive boson field $\phi(x)$\cite{sm}. That is,
\begin{eqnarray}
\ket{\Omega^{(s)}}=e^{is(L+K)}\ket{\Omega^{(0)}}\ .
\end{eqnarray}
$\ket{\Omega^{(s)}}$ is the vacuum state of Hamiltonian
\begin{eqnarray}
&&H^{(s)}=\int dk\:E_s(k)a_s^\dagger(k)a_s(k)\nn\\
&&E_s(k)=\sqrt{k^2+1}\sqrt{1+k^2 e^{2s}}\ 
\end{eqnarray}
annihilated by the annihilation operator at scale $e^{-s}$%$\ket{\Omega^{(s)}}$ is the vacuum state of Hamiltonian \cite{zou2019magic}\footnote{Note that our convention differs from \cite{zou2019magic} in the sign of $s$.}
%\begin{eqnarray}
%H^{(s)}(\Lambda)&=&\int \frac{dx}{2} \Big(\p_x\phi(x)^2+\pi(x)^2 \nn\\
%&& +\Lambda^2 e^{2s}\phi(x)^2+\frac{1}{\Lambda^2}(\p_x\pi(x))^2\Big)\nn
%\end{eqnarray}
%annihilated by the annihilation modes at scale $e^{-s}$. 
\footnote{We set the mass $m=\Lambda e^s=1$ and let the cut-off vary to exercise the RG flow of massive free boson field theory; see the supplementary material.} 
% \KF{We set the mass $m=\Lambda e^s=1$ and let the cut-off vary to exercise the RG flow of massive free boson field theory; see the supplementary material.}%Then, the annihilation operator is defined to kill $\ket{\Omega^{(s)}}$
: 
\begin{eqnarray}\label{generala}
&&a_s(k)=\sqrt{\frac{\alpha_s(k)}{2}}\phi(k)+\frac{i}{\sqrt{2\alpha_s(k)}}\pi(k)\nn\\
&&\alpha_s(k)=\sqrt{\frac{k^2+1}{k^2e^{2s}+1}}\ .
\end{eqnarray}
% They satisfy the standard commutation relations $[a_s^\dagger(k),a_s(k')]=\delta_{kk'}$.
The renormalized creation/annihilation operators are related to those of the UV theory 
\begin{eqnarray}
&&a_s^\dagger(k)\pm a_s(k) =\beta_s(k)^{\pm 1}(a^\dagger(k) \pm a(k))\nn\\
%+\frac{(\beta(k,s)-\beta(k,s)^{-1})}{2}a^\dagger(k),\nn\\
%&&\phi^s(k)=\beta(k;s)^{-1}\phi(k),\qquad \pi^s(k)=\beta(k;s)\pi(k)\nn\\
&&\beta_s(k)=\lb 1+k^2e^{2s}\rb^{1/4}\ .
\end{eqnarray}

\begin{figure}
    \centering
    \includegraphics[width=0.23\textwidth]{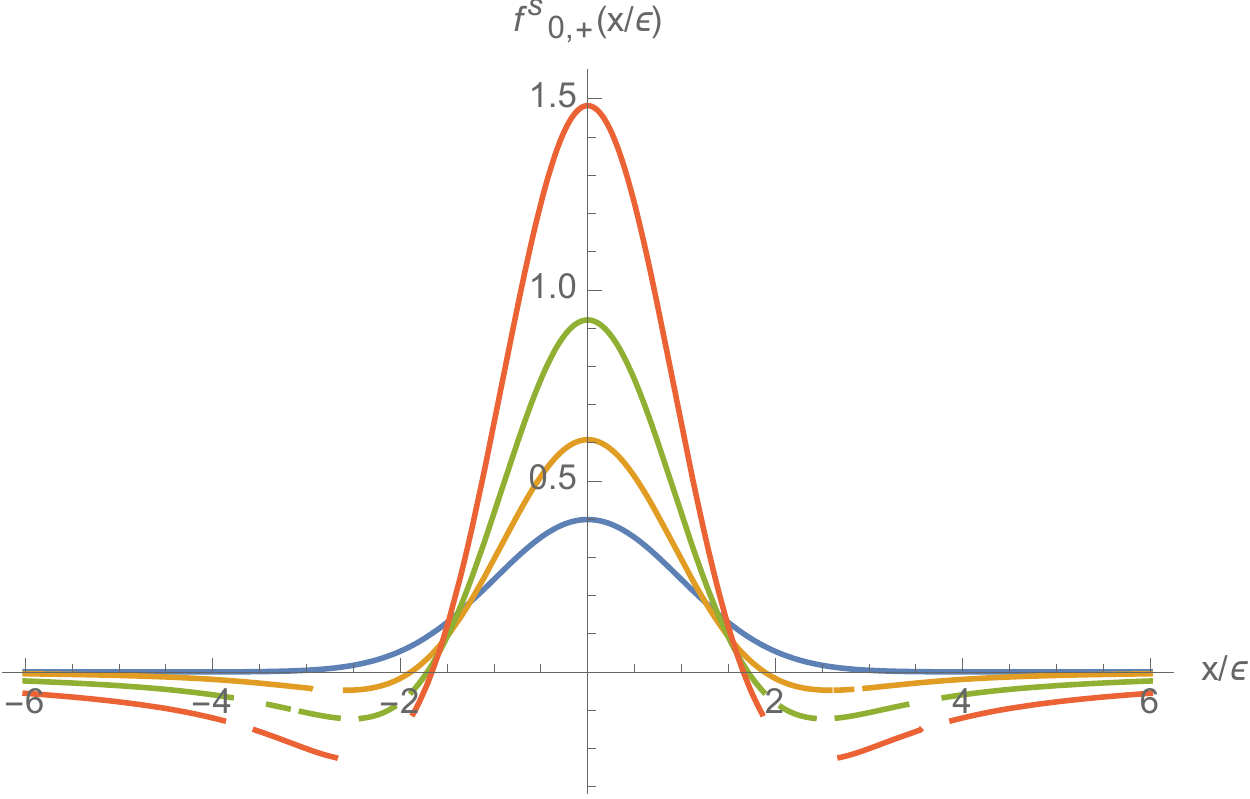}
    \includegraphics[width=0.23\textwidth]{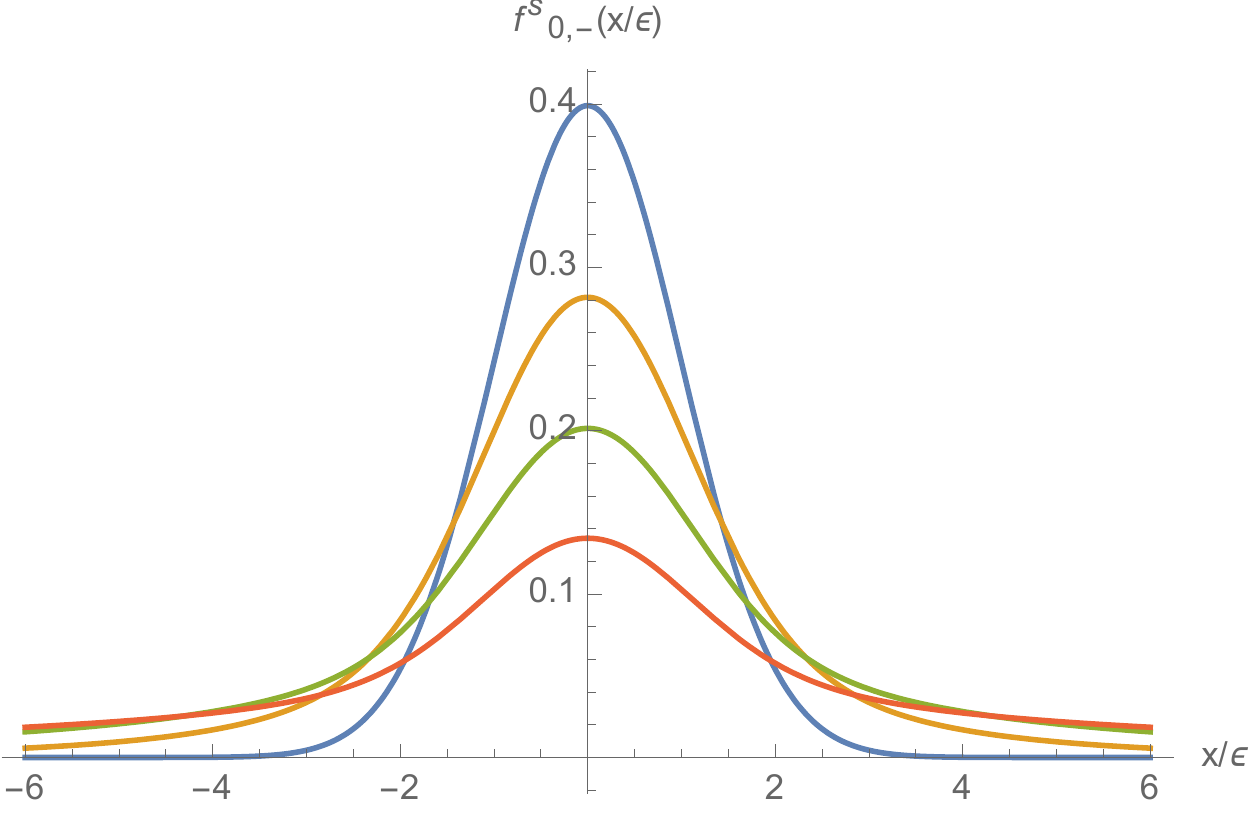}
    \caption{\small{The renormalization of the functions $f^s_{0,\pm}(x)$ is insignificant until the cut-off length scale becomes comparable to $\ep$ (width of $f_0(x)$). As we flow further towards the IR (Left) the function $f^s_{0,+}(x)$ becomes highly peaked (Right) the function $f^s_{0,-}(x)$ flattens out. (Blue: $s=-\infty$, Yellow: $s=1$, Green: $s=2$, Red: $s=3$.)}}
    \label{fig:numerical}
\end{figure}

We choose coherent operators as errors:
\begin{eqnarray}
&&D(f):=e^{a^\dagger(f)-a(f^*)},\qquad f=f_-+i f_+\nn\\
&&a^\dagger(f_-+if_+)-a(f_--if_+)=(a^\dagger-a)(f_-)+i(a^\dagger+a)(f_+)\nn
\end{eqnarray}
where $f_{\pm}$ are real functions \cite{sm}.
The renormalization of the coherent operator can be absorbed in the choice of smooth function $D(f)=D^{(s)}(f^s)
$ with \footnote{In real-space we have
$f^s_\pm(x)=B^{\pm 1/4}f_\pm(x)$ with $B:=(1-e^{2s}\p^2)$.} 
% \KF{In real-space we have
% $f^s_\pm(x)=B^{\pm 1/4}f_\pm(x)$ with $B:=(1-e^{2s}\p^2)$.}
\begin{eqnarray}\label{rencoherent}
&&f^s_\pm(k)=\beta_s(k)^{\pm 1}f_\pm(k),
\end{eqnarray}
%\qquad g^s_\pi(k)=\beta(k,s)^{-1}g_\pi(k)\nn\ .
Acting on the vacuum the coherent operator creates the coherent state with cut-off length $e^s$
\begin{eqnarray}
\ket{f;s}=D^{s}(f)\ket{\Omega^{(s)}}
\end{eqnarray}
% which has the energy
% \begin{eqnarray}
% &&\braket{f;s|H^{(s)}|f;s}=\int dk\: E_s(k) |f(k)|^2\ .
% % \nn\\
% % &&=(C^{1/4}f|C^{1/4}f),\nn\\
% % &&C=(\p^2+1)(1+e^{2s}\p^2)
% \end{eqnarray}
It is convenient to define the following inner product on the space of test functions
\begin{eqnarray}\label{testfunc}
&&(f|g):=\int dx f^*(x) g(x)\ .
\end{eqnarray}
% In general, knowing the behavior of smooth functions under a renormalization \footnote{We qualitatively discussed how a smooth function peaked at a point should behave under a renormalization.} is useful for the later analysis. 
Consider a function $f(x) = f_{-}(x) + if_{+}(x)$ with real smooth functions, $f_-$, $f_+$,  that is localized around $x=x_0$ with linear size $|A|$. Under the RG this function evolves to
\begin{eqnarray}\label{renf0}
&&f^s_{\pm}(x)=(1-e^{2s}\p_x^2)^{\pm 1/4}f_{\pm}(x)\ .
\end{eqnarray}
There are two distinct stages. When $e^s/|A| \ll 1$, the second term in $f^s$ above is negligible and the function is frozen $f^s(x) \approx f(x)$. The second stage starts at $e^s/|A| \approx 1$. For the function $f^s(x)$ deep in IR, both near the peak $|x-x_0|\ll |A|$ and far from the peak $|x-x_0| \gg |A|$, the second term $e^{2s}\p_x^2$ dominates:
\begin{eqnarray}\label{approx}
&&f^s_{\pm}(x)\simeq e^{\pm s/2}(\p_x^2)^{\pm 1/4}f(x)\ .
%   \nn\\
%   &&f^s_{0,+}(x)=\frac{e^{s/2}e^{-\tilde{x}^2}}{\sqrt{2\ep^3|\tilde{x}|}}\lb \tilde{x}^2I_{5/4}(\tilde{x}^2)-(\tilde{x}^2-1/2)I_{1/4}(\tilde{x}^2)\rb\nn\\
%   &&f^s_{0,-}(x)=\frac{e^{-s/2}e^{-\tilde{x}^2}\sqrt{|\tilde{x}|}}{\sqrt{2\ep}|\tilde{x}|}I_{-1/4}(\tilde{x}^2)
\end{eqnarray}
%   where $\tilde{x}=x/\ep$. 
The function $f^s_{+}$ ($f^s_{-})$ grows (decays) exponentially fast as $e^{(s-\log|A|)/2}$ $(e^{-(s-\log|A|)/2})$ in the IR, respectively; see (Fig. \ref{fig:numerical})\footnote{For this calculation, we choose a Gaussian wave-packet
\begin{eqnarray}\label{Gaussian}
f_0(x)=\frac{1}{\sqrt{2\pi}\ep}e^{-\frac{(x-x_0)^2}{2\ep^2}}
\end{eqnarray}
for $f_0(x)$}. 
  
% \KF{The renormalization of the functions $f^s_{0,\pm}(x)$ is insignificant until the cut-off length scale becomes comparable to $\ep$ (width of $f_0(x)$). As we flow further towards the IR (Left) the function $f^s_{0,+}(x)$ becomes highly peaked (Right) the function $f^s_{0,-}(x)$ flattens out. (Blue: $s=-\infty$, Yellow: $s=1$, Green: $s=2$, Red: $s=3$.)}

Note that the larger $|A|$ is, the later the second stage starts. Because of the above behaviour, for a coherent operator $D(f)$ as an error operator with a function $f$ supported on a large $|A|$, one needs to flow much deeper in the IR to achieve the  error-correction code with the same errors.

%Add more comments about the role of the second term, and intuitive comparison with MERA!!

%near the peak $|x-x_0|\ll |A|$, . On the contrary, it behaves ... far from the one $|x-x_0| \gg |A|$.

%add approximation.

%we are interested in its behaviour near the peak $|x-x_0|\ll |A|$ and far from the one $|x-x_0| \gg |A|$ since they are the key to make the encoding work and derive the error-correction condition.

\subsection{Encoding a qudit at a point:}\label{encoding}

%then, the later, we use gaussian for the actual example and calculation

%by the way, if one can correct the error $D(ipf)$, the codes are too much. The errors should not map one code state to the other. Otherwise, one cannot notice which one was the initial code state. This results in the impossibility of recovery. There might be something I am missing.

We use a coherent operator $D(ipf_0)$ with a real function $f_0$ and $p$ running from $0$ to $q-1$, i.e.
\begin{eqnarray}
\ket{p,x_0}:=D(ipf_0)\ket{\Omega}\ .
\end{eqnarray}
to encode a $q$-level quantum system at $x=x_0$.
The set of states $\ket{p,x_0}$ are almost orthonormal to each other for large $(f_0|f_0)$ because 
\begin{eqnarray}
\braket{p',x_0|p,x_0}&=&\braket{D(i(p-p')f_0)}=e^{-\frac{1}{2}(p-p')^2(f_0|f_0)}\simeq\delta_{pp'}\nn.
\end{eqnarray}
The logical algebra of this encoding is generated by two operators $D(iqf_0)$ and $D(g_-)$ for a smooth real function $g_-$. The operator $D(iqf_0)$ takes us in between code states
\begin{eqnarray}\label{logicalmom}
\braket{p',x_0|D(i qf_0)|p,x_0}=\delta_{p',p+q}
\end{eqnarray}
and the operator $D(g_-)$ is diagonal in the basis $\ket{p,x_0}$:
\begin{eqnarray}
\braket{p',x_0|D(g_-)|p,x_0}&=&\delta_{pp'}\braket{D(g_-^s)}e^{-i(p+p')(g_-|f_0)}\nn\ .
%&=&\delta_{pp'}e^{-i(p+p')q}\ .
\end{eqnarray}
%where $(g_-|f_0)\simeq g_-(x=x_0)$ in the $\ep\to 0$ limit.
If $P_0$ is the projection to the code subspace spanned by $\ket{p,x_0}$ then the operators $P_0$, $P_0 D(iqf_0)P_0$ and $P_0D(g_-)P_0$ and their Hermitian conjugates generate the algebra of the $q$-level system encoded at point $x=x_0$.

%The above code subspace is at UV. The IR code subspace consists of the code states defined by 
%\begin{equation}
%    \ket{p,x_0;s}=D^s(ipf_0) \ket{\Omega^{(s)}}.
%\end{equation}
%It is simple to check their approximate orthogonality
%\begin{eqnarray}
%\braket{p',x_0;s|p,x_0;s}&=&\braket{D^s(i(p-p')f_0)}=??check
%? e^{-\frac{1}{2}(p-p')^2(f_0|f_0)}\simeq\delta_{pp'}\nn.
%\end{eqnarray}

\subsection{Error correction condition:}

%Here, we discuss that any set of UV logical coherent operators $D(h_-)$, $D(ih_+)$ for real smooth functions $h_-$, $h_+$ supported either A,C, or both are correctable if we go deep enough in the IR. In order to distinguish the smooth function for the error operators from the one used for the encoding, we use $h_-$ and $h_+$ instead of $g_-$ and $f_0$. To show the validity of our statement, we need to argue that $\braket{p',x_0;s|D(h_-)|p,x_0;s}$ and $\braket{p',x_0;s|D(ih_+)|p,x_0;s}$ satisfy the condition in (\ref{eq:klcond}).

The Knill-Laflamme condition for approximate quantum error operator \cite{beny2010general} tells us that we can approximately correct for error caused by the operator $\mO$ if and only if this operator is proportional to the projection to the code subspace up to small corrections $\delta$:
\begin{eqnarray}
\braket{\Psi_r|\mO|\Psi_{r'}}=c \delta_{rr'}+\delta\ .
\end{eqnarray}
We choose the coherent state with cut-off length scale $e^{s}$ as the states of the code subspace and choose as errors the UV logical coherent operator $D(g_-)$ and $D(iq f_0)$. To show error correction we need 
\begin{eqnarray}
&&\braket{p',x_0;s|D(g_-)|p,x_0;s}=c(g_-,s)\delta_{pp'}+\delta_1\nn\\
&&\braket{p',x_0;s|D(iqf_0)|p,x_0;s}=c(f_0,s)\delta_{p,p'}+\delta_2
\end{eqnarray}
for some functions $c(g_-,s)$ and $c(f_0,s)$ and small $\delta_1$ and $\delta_2$.
In the supplementary material, we compute these matrix elements and find that approximate error correction conditions above are satisfied for any fixed set of UV logical operators $D(h)$ with $h$ supported on $A$, $C$ or both if we go deep enough in the IR. 
The operators supported on $A$ are correctable when $e^s\gg |A|$ and those on $C$ are correctable when $C$ is far enough from $x=x_0$. In fact, we can consider the operators that are supported on $AC$ and the same argument above implies that deep in the IR the UV operators of $AC$ can be corrected. This is reminiscent of uber-holography and local error correction in \cite{Pastawski:2016qrs,kim2017entanglement}. Note that we are not correcting for erasures. For any fixed IR scale $e^s$ there always exist $D(h)$ supported on $A$ or $C$ with $(h|h)$ large enough that can distinguish the code states, however such coherent operators carry enormous amounts of energy.

%As opposed to the MERA where we were protected against the erasure of a region, in cMERA we have the weaker statement that given a set of local coherent operator we are protected from them if we go deep enough in the IR. 

%To work this out explicitly, in \cite{sm}, for encoding, 

% \paragraph{Adding interactions}

% Consider the Wilsonian one-loop ground-state wave-function of scalar field theory with $\lambda \phi^4$ interaction. We can engineer the unitary flow (cMERA) that evolves the state as one desires. This was done in \cite{cotler2019renormalization}. \NL{Look at the massive answer and the $\lambda$-mixing term.}

\section{Holographic RG and error correction} \label{sec:holography}

%summarise the problem and statement

% Trade-off bound is one way to characterize the properties of quantum error-correction code[]. The well-known trade-off bound for a quantum code is called quanutm singleton bound. However, it is sensible only for finite dimensional system since it involves the dimension of the Hilbert space of a whole system. 
Holography can be viewed as a QEC where the algebra of bulk regions are encoded on the boundary regions such that they are protected against local boundary erasures \cite{Almheiri:2014lwa}.
To characterize the properties of such holographic QEC, the authors of \cite{Pastawski:2016qrs} defined the notions of price $p_{X}$ and distance $d_{X}$ of the logical algebra associated to a bulk region $X$ that we review in the supplementary material. The holographic Singleton bound is the statement that difference of price and distance for any $X$ can not be less than the number of logical degrees of freedom $k_{X}$ in that bulk region \cite{Pastawski:2016qrs}:
\begin{align}
    k_{X} \, \le \, p_{X} - d_{X} \, .\label{eq-holo-singleton}
\end{align}

In the following subsection, we consider a simple example of a holographic RG flow in which there exists finite volume regions of the bulk $X$ that satisfy $p_X=d_X$. The holographic Singleton bound then implies that the rather unphysical result that there should not be any logical degrees of freedom in that subregion. In Sec.~(\ref{eq-sec-rw}), we resolve this problem by pointing out that near phase transition points we needs to use a modified definitions of distance and price that is formulated using the reconstructable wedge instead of the entanglement wedge.

\subsection{Violation of holographic strong Singleton bound and Hologrpahic RG flow}

%state the problem and its cause concisely. 

Suppose we set off an RG flow on the CFT by deforming with a relevant operator. The geometry that is dual to the RG flow on the boundary is given by \cite{Girardello:1998pd,Freedman:1999gp,Girardello:1999bd}
\begin{align}
    ds^{2} \, = \, e^{2 A(r)} \, \left( -dt^{2} + dx^{2} \right) \, + \, dr^{2} \, ,
\end{align}
where $A(r)$ is such that $A(r) \, \sim \, {r/L_{UV}}$ near $r = \infty$ whereas $A(r) \, \sim \, {r/L_{IR}}$ near $r = - \infty$. $L_{UV}$ is related to the central charge of the UV theory %according to Eq.~\eqref{eq-cl-rel}
, $c = 3L/2G_{N}$, whereas $L_{IR}$ is related to the central charge of the IR theory to which the UV theory flows \footnote{The holographic $c$-theorems say that the null energy condition in the bulk implies $L_{IR} < L_{UV}$ \cite{Myers:2010xs,Myers:2010tj}.}.
% \KF{The holographic $c$-theorems say that the null energy condition in the bulk implies $L_{IR} < L_{UV}$ \cite{Myers:2010xs,Myers:2010tj}.} 
The function $A(r)$ captures the flow of the boundary theory from UV to IR \cite{Myers:2010xs,Myers:2010tj}. 

In this work, we consider a simple example where $A(r)$ is given by a
\begin{align}
    A(r) \, = \, 
    \begin{cases}
        r/L_{UV} & \quad r \geq 0 \\
        r/L_{IR} & \quad   r \leq 0 
    \end{cases} \, .
\end{align}
\begin{figure}
    \centering
    \includegraphics[width=0.35\textwidth]{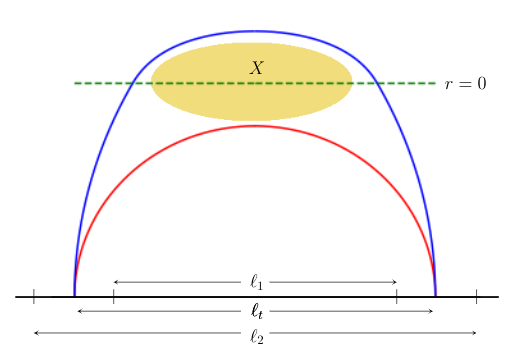}
    \caption{\small{Pictorial representation of the phase transition in the HRT surfaces at $\ell = \ell_{t}$ for $L_{IR} = 0.3$ and $L_{UV} = 1.0$. The critical length scale is in the range $\ell_{1} < \ell_{t} < \ell_{2}$. The HRT surface for $\ell$ just bigger/smaller than $\ell_{t}$ is shown in blue/red color. The yellow shaded region is an example of a finite size region $X$ for which the distance and price are equal.}}
    \label{fig-jump}
\end{figure}
This simple model of holographic RG flow with the extremal area surfaces corresponding to boundary regions was studied in \cite{Myers:2012ed,Albash:2011nq}. In \cite{Myers:2012ed}, there are three characteristic single intervals, $\ell_1$, $\ell_2$, $\ell_t$ on the boundary denoted in fig.\ref{fig-jump}. For any HRT surfaces anchored on a boundary subregion whose interval $\ell$ satisfy $\ell < \ell_1$, the corresponding HRT surface is always placed within the bulk UV region. On the contrary, when $\ell>\ell_2$, a HRT surface penetrate into the bulk IR region. When $\ell_1<\ell < \ell_2 $, there is a possibility that HRT surface could be in either in the IR or the UV region. That is, there exists a critical size of the interval, $\ell_{t}$, below which the surfaces that stay in the UV region has a smaller area and above which the surfaces that reaches the IR region has a smaller area. \footnote{In \cite{sm}, we numerically calculate the critical size of the interval by comparing the area of the surfaces involved. In addition, the result of this analysis is shown as a plot of $\ell_{t}$ versus $L_{IR}/L_{UV}$.}.
% \KF{In \cite{sm}, we numerically calculate the critical size of the interval by comparing the area of the surfaces involved. In addition, the result of this analysis is shown as a plot of $\ell_{t}$ versus $L_{IR}/L_{UV}$.} 
Due to the `phase transition' in the HRT surface at $\ell = \ell_{t}$, there is a jump in the bulk entanglement wedge as well in the same way\footnote{The jump in the entanglement wedge at $\ell = \ell_{t}$ can be measured in terms of the proper distance between these minimum radial points. We numerically calculate this proper distance in \cite{sm} and find that we can make this proper distance bigger by making the difference between $L_{IR}$ and $L_{UV}$ bigger; see %(Fig.~\ref{fig-holo-rg} Right)
in \cite{sm}.}.
% \KF{The jump in the entanglement wedge at $\ell = \ell_{t}$ can be measured in terms of the proper distance between these minimum radial points. We numerically calculate this proper distance in \cite{sm} and find that we can make this proper distance bigger by making the difference between $L_{IR}$ and $L_{UV}$ bigger; see %(Fig.~\ref{fig-holo-rg} Right)
% in \cite{sm}.}

The transition in the entanglement wedge at $\ell = \ell_{t}$ has interesting implications for holographic QECC. Consider a bulk region $X$ which is the intersection of the region $r < r_{UV} $ and the entanglement wedge of a boundary interval of size slightly greater than $\ell_{t}$. The condition $r < r_{UV} $ implies that the region $X$ is not in the entanglement wedge of any boundary interval of size less than $\ell_{t}$. Hence, according to the definitions from \cite{Pastawski:2016qrs}, the distance of the logical algebra associated to region $X$ is given by $\ell_{t}$ as well as the price. %Eqs.~\eqref{eq-holo-d-point}-\eqref{eq-holo-d},the distance of the logical algebra associated to region $X$ is given by $\ell_{t}$. Moreover, according to Eq.~\eqref{eq-holo-p}, the price of the region $X$ is also given by $\ell_{t}$\footnote{The distance and the price are actually equal to $(\ell_{t})^{\alpha}$ where $\alpha = \log(2)/\log(\sqrt{2}+1)$ \cite{Pastawski:2016qrs}. This is the size of the fractal like disconnected intervals such that the entanglement wedge of the disconnected region has the same minimum radial point as the entanglement wedge of a single interval of size $\ell_{t}$. This construnction is called \textit{uberholography} in \cite{Pastawski:2016qrs}.}. 
This means that we have found a logical subalgebra associated to a finite volume bulk region for which the price and the distance are the same. 
Comparing this with the holographic strong Singleton bound in Eq.~\eqref{eq-holo-singleton}, we deduce the surprising conclusion that the number of logical degrees of freedom in that finite volume subregion should be zero. In the next subsection, we argue that to resolve this seeming paradox one has to modify the definition of the distance and the price using the concept of reconstruction wedge \cite{Akers:2019wxj}. 
% or else we get a violation of the holographic strong Singleton bound.  

% We discuss in the next subsection how to modify the definition of the distance and the price to resolve this apparent paradox.

\subsection{Price, distance, and the reconstruction wedge} \label{eq-sec-rw}

%solution: redefine the bound with RW. how?

% We observed in the previous subsection that a phase transition in the entanglement wedge led us to a violation of the holographic strong Singleton bound. This violation can be resolved by introducing the concept of reconstruction wedge\cite{Akers:2019wxj}.

The idea of reconstruction wedge is motivated by the observation that the entanglement wedge can be different for different states in the code subspace
\cite{Hayden:2018khn,Akers:2019wxj,Akers:2020pmf,wang2021refined}. %In particular, it was argued in \cite{Akers:2019wxj} that the bulk region that can be reconstructed given a boundary region $B$ is not the entanglement wedge of $B$. In fact, this region can be macroscopically smaller than the entanglement wedge, $\mathcal{E}(B)$. 
The \textit{reconstruction wedge}, $\mathcal{R}(B)$, corresponding to the boundary region $B$ is defined to be the intersection of all the entanglement wedges of $B$ for every state in the code subspace \cite{Akers:2019wxj} . 
% Hence, the reconstruction wedge is macroscopically smaller than the entanglement wedge.

The fact that the reconstruction wedge is smaller than the entanglement wedge of any state of that code subspace is the key to resolve the paradox above. The price of a logical subalgebra is the smallest boundary region on which any logical operator can be represented.
Similarly, the distance of a logical subalgebra is the smallest boundary region $B$ such that the logical algebra cannot be reconstructed from the complement boundary region, $B^{c}$. 
It is natural to define distance and price in terms of the reconstruction wedge rather than the entanglement wedge. We propose the following definitions of the price and distance for a logical algebra associated to a bulk region $X$: %modified from Eq.~\eqref{eq-holo-p} to
\begin{align}\label{eq-holo-p-R}
    &p_{X} \, = \, \min_{B: X \in \mathcal{R}(B)} \, |B| \,\nn\qquad  
%\end{align}
% Similarly, the distance of a logical subalgebra is the smallest boundary region $B$ such that the logical algebra cannot be reconstructed from the complement boundary region, $B^{c}$. Similarly, it is more natural to define the distance in terms of the reconstruction wedge. Therefore, we propose that the distance of a subregion $X$ is
%\begin{align}
    d_{X} \, = \, \min_{x \in X} \, d_{x} \, ;  \quad\quad\quad\quad\quad \\
    &d_{x} \, = \, \min_{B: x \notin \mathcal{R}(B^{c})} |B| \, .
\end{align}

The modifications above resolve the seeming violation of the holographic Singleton bound. For the region $X$ that we defined, the distance is still determined by the $\ell = \ell_{t}$. The price, on the other hand, is determined by $\ell = \ell_{2}$ which is the largest length for which the surface % in Eq.~\eqref{eq-surf-uv}
exists\footnote{Again, the distance and the price are given by $(\ell_{t})^{\alpha}$ and $(\ell_{2})^{\alpha}$ respectively where $\alpha = \log(2)/\log(\sqrt{2}+1)$ as determined by the uberholography construction.}. 
% \KF{Again, the distance and the price are given by $(\ell_{t})^{\alpha}$ and $(\ell_{2})^{\alpha}$ respectively where $\alpha = \log(2)/\log(\sqrt{2}+1)$ as determined by the uberholography construction.}

\section{Summary and discussions}\label{sec:conclude}

In summary, in this work, we further developed the connection between RG and approximate error correction codes by providing the following two examples: 1) the RG flow of classical Ising model as a classical code 2) continuous MERA for massive free fields as a quantum code. We also considered holographic RG flows for a two dimensional boundary theory and argued that the phase transition in the entanglement wedge points to the fact that we need to define the notions of price and distance using the reconstructable wedge.

In this work, we postulated the picture that the Hilbert space of an effective field theory with the cut-off scale $\Lambda$ should be viewed as a code subspace of all states that are approximately protected against the short-distance errors localized on a region of linear size $A$ much smaller than the cutoff, $|A|\lesssim 1/\Lambda$. To argue for this point, we used cMERA as a concrete realization of the real-space RG flow of massive free fields. However, there are other approaches to the RG flow. Examples include the continuous Tensor Network Renormalization (cTNR) in \cite{hu2018continuous}, the generalization of cMERA using Euclidean path-integrals \cite{nozaki2012holographic,miyaji2015boundary}, the RG flow for free $O(N)$ model using Polchinski's exact RG \cite{fliss2017unitary}.  In some of these approaches the map from the IR physics to the UV is no longer an exact isometry. It is an interesting question to investigate the approximate QECC code appears in these other approaches to the RG flow.

\section*{Acknowledgements}

NL would like to thank the Institute for Advance Study for their
hopsitality and the NSF grant PHY-1911298. We thank Venkatesa Chandrasekaran, Nicholas Laracuente and Shoy Ouseph. NL is very grateful to the DOE that supported this work through grant DE-SC0007884 and the QuantiSED Fermilab consortium.

%\appendix

% \section{Wick's theorem}

% The form of the Wick theorem that is 

\bibliography{main}

\end{document}

\preprint{APS/123-QED}

\title{Supplemental Material: Renormalization group and approximate error correction}% Force line breaks with \\
\author{Keiichiro Furuya$^{1}$, Nima Lashkari$^{1,2}$, Mudassir Moosa$^{1}$}
\affiliation{
 ${}^1$ Department of Physics and Astronomy, Purdue University, West Lafayette, IN 47907, USA}
% \author{Nima Lashkari}%
%  \email{lashkari@purdue.edu}
%  \altaffiliation[Also at ]{School of Natural Sciences, Institute for Advanced Study, Princeton, New Jersey 08540, USA}
 %


% \author{Mudassir Moosa}
%  \homepage{http://www.Second.institution.edu/~Charlie.Author}
\affiliation{
  ${}^2$ School of Natural Sciences, Institute for Advanced Study, Princeton, New Jersey 08540, USA}%
% \affiliation{
%  Third institution, the second for Charlie Author
% }%
% \author{Mudassir Moosa}
% \affiliation{%
%  Authors' institution and/or address\\
%  This line break forced with \textbackslash\textbackslash
% }%

% \collaboration{CLEO Collaboration}%\noaffiliation

%              %  but any date may be explicitly specified

%\begin{abstract}

%In renormalization group (RG) flow, the low energy states form a code subspace that is approximately protected against the local short-distance errors. We motivate this connection with an example of spin-blocking RG in classical spin models. We consider the continuous multi-scale renormalization ansatz (cMERA) for massive free fields as a concrete example of real-space RG in quantum field theory (QFT) and show that the low-energy coherent states are approximately protected from the errors caused by the high-energy localized coherent operators. In holographic RG flows, we study the phase transition in the entanglement wedge of a single region and argue that one needs to define the price and the distance of the code with respect to the reconstructable wedge.
% The trade-off bounds set a bound on the amount of quantum information at a scale. We comment on the connections to the quantum error correction codes in holography and the high-energy states of chaotic quantum systems.
%\end{abstract}

\maketitle

In section \ref{sec:rglqecc}, we briefly review the recent work on the local quantum error-correction in MERA. In section \ref{sec:classicalspinblock}, we describe the classical and quantum spin blocking, and discuss how they are related to quantum error-correction. In section \ref{sec:cMERA}, we review the cMERA construction in \cite{zou2019magic} for massive free boson field theory. In section \ref{sec:RGcoherent}, we define our error model and describe the RG flow of the error operators. In section \ref{encoding}, we check the Knill-Laflamme approximate QEC condition for the cMERA.
% It allows us to construct the logical algebra in the UV regime. %It shows that the encoding prepares the code subspace satisfying the essential properties,
At last, in section \ref{sec:holography}, we review the properties of holographic QEC and the reconstruction problem. We provide the details of the holographic phase transition example discussed in the text.

\section{Renormalization group and Local quantum error-correction} \label{sec:rglqecc}

There are three main possibilities for the IR end points of an RG flow: 1) mass gap 2) massless particles 3) a scale-invariant fixed point. Examples of each type are the following: Quantum Chromodynamics (QCD) is type one because under the RG the coupling grows and we end up with massive glueballs in the infrared. Quantum Electrodynamics (QED) and models with spontaneous breaking of a continuous symmetry are of second type. Photons and the Nambu-Goldstone boson are the surviving massless mode in the IR. Non-Abelian Yang-Mills theory coupled to fermions in the conformal window flows to the Bank-Zach fixed point in the IR, and the $\lambda \phi^4$ scalar theory in $D=4-\epsilon$ dimensions flows to the Wilson-Fisher fixed points. %These are both examples of type three.

The connection between RG and quantum error-correction becomes clear when
 the groundstate is degenerate due to either symmetry breaking or the existence of topological order. For instance, the two-dimensional Ising model at low temperature breaks the $\mathbb{Z}_2$ symmetry spontaneously by forming long-range ordered ferromagnetic states $\ket{00\cdots 0}$ and $\ket{1\cdots 1}$. This is a classical repetition code that corrects for local bit flips ($\sigma_X$ Pauli matrix). The existence of a local order parameter that distinguishes different code states precisely implies that the local density matrices are distinguishable and one cannot correct local quantum errors ($\sigma_z$ Pauli matrix). 

We obtain a QEC in degenerate ground subspace if there is topological order.
Topologically ordered systems such as the Toric code can be used to encode quantum information such that it is protected against local errors.

\begin{figure}
    \centering
    \includegraphics[width=0.4\linewidth]{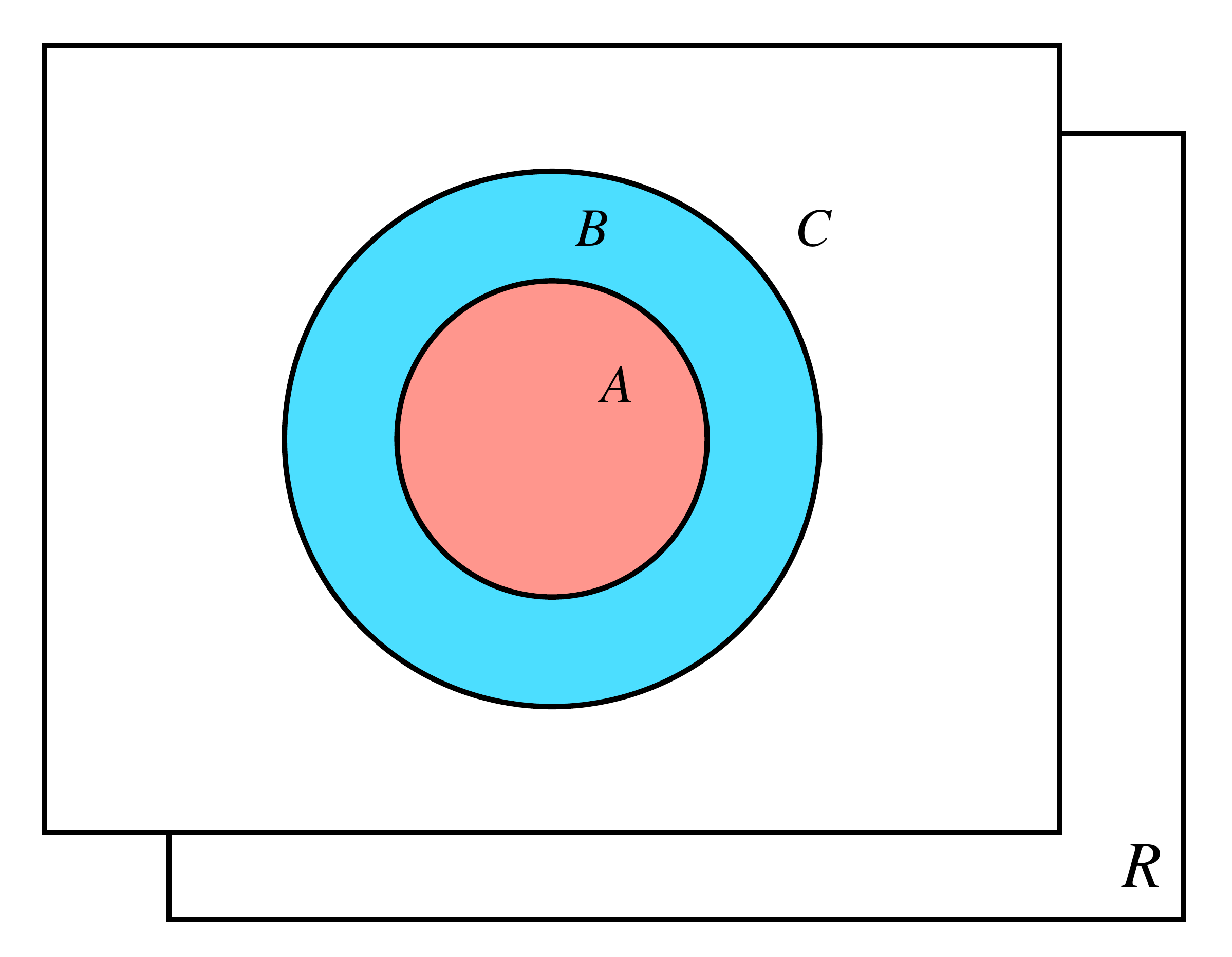}
    \caption{\small{The spatial region is partitioned by $ABC$ with purifying system $R$. The local erasure $tr_A$ acts on the red-colored region A. The blue-colored region is the spatial domain of a local recovery ma, $\mR_{B}^{AB}$.}}
    \label{fig1}
\end{figure}

%If we call the subspace spanned by the ground states the code subspace and purify it in a reference system $R$, the above definition is equivalent to the decoupling condition $I_\rho(A:R)=0$, which is the definition of an exact erasure-correction quantum code.
Following \cite{flammia2017limits} we consider local quantum error correction (LQEC) codes; see figure \ref{fig1} A QEC is called local if its stabilizers or gauge generators are supported on a small bounded region of the space.
%A QEC is called local if its stabilizers or gauge generators are supported on a small bounded region of the space.
% There is actually a case which LQECC arises from a certain topological ordered system. Here, we briefly review it as well as LQECC. 
On a lattice, LQECC are represented by four parameters $[[n,k,d]]_q$, where at each site of the lattice we have a $q$-level system, $n$ is the number of physical sites,  $k$ is number of the logical ``$q$-bits'', and $d$ is the distance of the code. The distance is defined to be size of the support of the smallest logical operation. In QEC, there is a trade-off between the rate of logical q-dits we encode $k/n$ and the distance $d$. For a fixed rate the distance has an upper bound. For instance, for local commuting codes in $D$-spatial dimensions we have the trade-off bound $k d^{2/(D-1)}\leq c n$ where $c$ is a constant \cite{bravyi2010tradeoffs,haah2012logical}. A subclass of LQEC codes constructed out of local commuting projections includes the topological ordered systems such as Kitaev's quantum double model and the Levin-Wen string-nets \cite{kitaev2003fault,levin2005string,cui2020kitaev}. In continuum topological quantum field theories, it is expected that the states prepared using Euclidean path-integral lead to QEC codes \cite{salton2017entanglement}. In the case of Abelian Chern-Simons theory, the ground subspace is a stabilizer code. If the topological order is stable small perturbations of commuting projection codes leads to approximate error correction codes \cite{flammia2017limits}.

\section{Spin-Blocking}

\subsection{Classical}\label{sec:classicalspinblock}

\begin{figure}
    \centering
    \includegraphics[width=0.6\linewidth]{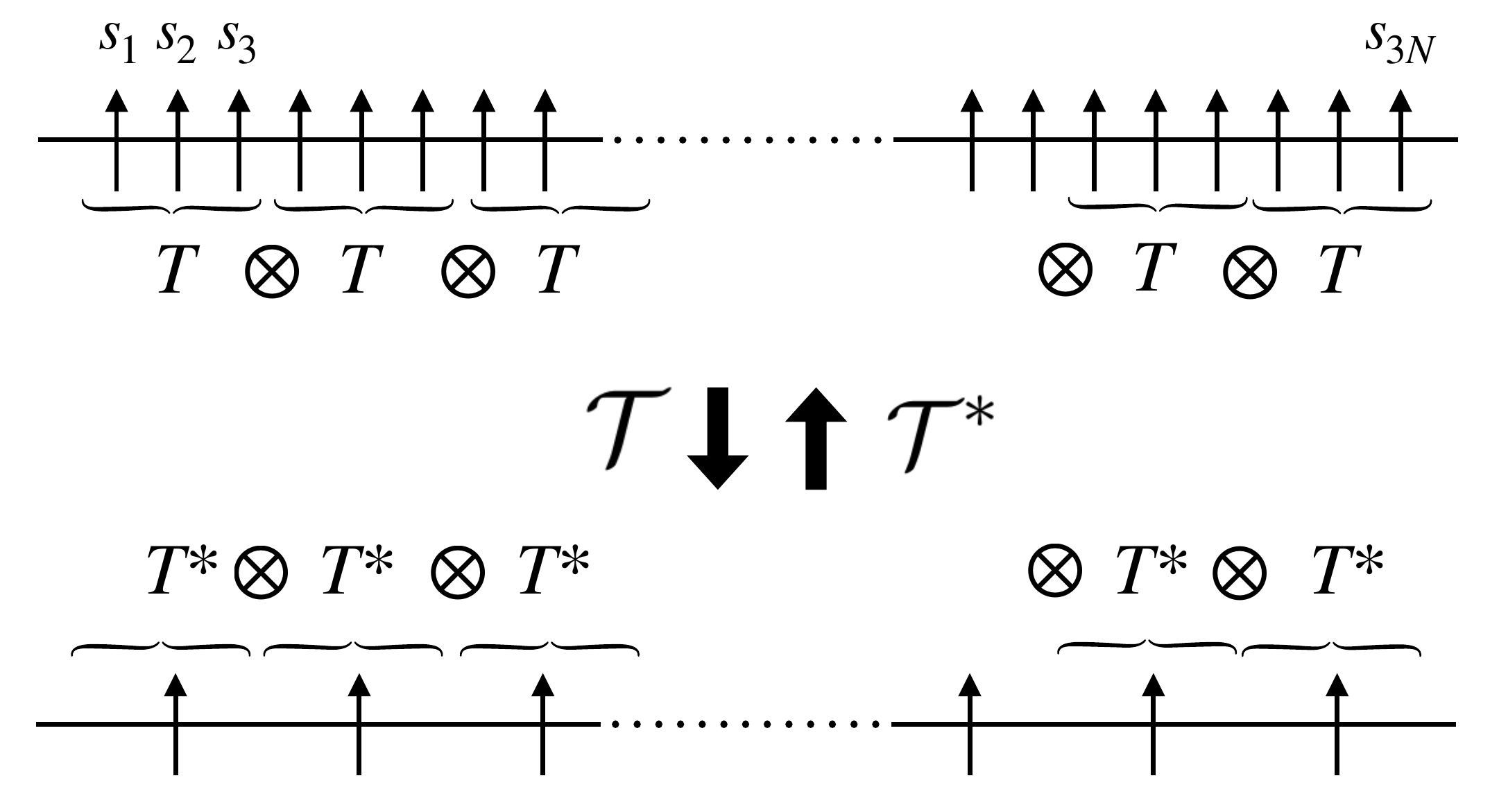}
    \caption{\small{This is a single step coarse-graining on a classical Ising model in one dimension with $3N$ sites. Since $T$ coarse-grains three spins into a single spin, $\mT = T^{\otimes N} $ coarse-grains $3N$ sites into $N$ sites.}}
   \label{fig2}
\end{figure}

%In this section, we worked out the detailed construction of approximate classical code using a one-dimensional Ising model. %In classical physics, the standard real-space RG scheme of a spin systems involves splitting the lattice into blocks and coarse-graining the degrees of freedom localized inside each block. 
%The coarse-graining replaces the collection of spins inside the block $\{s_i\}$ with a collective degree of freedom $s'$ \cite{kadanoff1967static,wilson1975renormalization}. For example, 
In the main text, we discussed the construction of a classical code using a one-dimensional Ising model \footnote{The ground states of the one-dimensional nearest neighbor Ising model form an exact repetition code. A simultaneous $\mathbb{Z}_2$ flip on all sites is a logical operation which takes us from one code state to another in the repetition code.}. In this section, we supplement the details of model for the completeness. The model we consider is the translation-invariant one dimensional classical Ising model  with $3N$ sites, a local Hamiltonian
\begin{eqnarray}
H(\{s\})=-J\sum_{i=1}^{3N-1} s_i s_{i+1}+h\sum_{i}^{3N-1} s_{i}
\end{eqnarray}
and periodic boundary conditions; see figure \ref{fig2}.
%where $s_i$ is the $i^{th}$ spin variable and $i=1,\cdots, 3N$.
The configuration space is the set of all sequences  $\{s\}=\{s_1s_2\cdots s_{3N}\}$ and the thermal state is the probability distribution
\begin{eqnarray}
p(\{s\})=e^{-\beta H(\{s\})}/Z
\end{eqnarray}
on this configuration space. 

Here, we represents states of first three spins as vectors $\ket{s_1s_2s_3}$, and the coarse-graining map as a matrix $T$ describing a classical information information channel acting on probability vectors. For example, the matrix elements are $\braket{s'|T|s_1s_2s_3}$. In our example, the map $T$ is $T=\big(\begin{smallmatrix}
  1 &1&1&0&1&0&0&0\\
  0 &0&0&1&0&1&1&1 
\end{smallmatrix}\big)$. In general, applying the one step coarse-graining to the whole system can be written as $T^{\otimes N} \ket{s_1s_2s_3} \otimes \cdots \otimes \ket{s_{3N-2}s_{3N-1}s_{3N}} $. As mentioned in the main text, we think of the normalized transpose map $T^*\equiv T^T/4$ as an encoding isometry, the $s'$ variable as the logical state and the $\{s\}$ variables as the physical states. %A simple coarse-graining scheme is the majority vote scheme
%\begin{eqnarray}\label{majvote}
%s'=\begin{cases}
%    +1       & \quad \text{if} \: s_1+s_2+s_3\geq 0\\
%    -1  & \quad \text{otherwise}\ .
%    \end{cases}
%\end{eqnarray}
%It is convenient to represent states as vectors $\ket{s_1s_2s_3}$ and think of the coarse-graining map as a matrix $T$ describing a classical information information channel acting on probability vectors \footnote{The matrix elements are $\braket{s'|T|s_1s_2s_3}$. In our example, the map $T$ is $T=\big(\begin{smallmatrix}
  %1 &1&1&0&1&0&0&0\\
  %0 &0&0&1&0&1&1&1 
%\end{smallmatrix}\big)$.}. 
%We think of the normalized transpose map $T^*\equiv T^T/4$ as an encoding isometry, the $s'$ variable as the logical state and the $\{s\}$ variables as the physical states. 
Hence, the code subspace is spanned by the two states $\ket{T^*(\pm 1)}$. In particular, the encoded states are
$\ket{T^*(\pm 1)}=\frac{1}{4}(\ket{\pm1,\pm1,\pm1}+\ket{\pm1,\pm1,\mp1}+\ket{\pm1,\mp1,\pm1}+\ket{\mp1,\pm1,\pm1})$.

\subsection{MERA}\label{sec:MERA}

%The intuitive discussion above generalizes to the renormalization group flow of quantum systems with local Hamiltonians. 
%In a gapped system, the RG flow becomes trivial at scales above the correlation length. Since we are interested in repeating the RG map many times, we focus on the real-space RG in critical systems. 
MERA is a quantum version of classical spin-blocking discussed above. % We start with a lattice theory in the IR. 
Consider a lattice theory in the IR. %The RG map can be viewed as an encoding isometry $W:\mH_{IR}\to \mH_{UV}$. 
In MERA, an encoding isometry $W:\mH_{IR}\to \mH_{UV}$\footnote{In the general theory of error correction, it suffices to take $W$ to be an approximate isometry. In continuum theories it can be unitary.} corresponds to two layers, first a layer of local isometries $V\otimes \cdots \otimes V$, and second a layer of local unitaries $U\otimes \cdots \otimes U$ called the disentanglers. %The layer of local isometries is the quantum analog of $\mathcal{T}^*$ map above.
In real-space RG, the disentanglers are essential to correctly remove the UV entanglement. By the same logic as in the classical case, we view the RG flow as a quantum channel  $\Phi(\mO_{UV})=W^{\dagger}\mO_{UV} W$ acting on UV operators $\mO_{UV}$. The RG flow of the operator has two stages: first, the size of its support shrinks monotonically. Once it is supported on just a single site, the second stage starts where the effect of the operator gets exponentially weaker, where the exponent is $\nu=-\log (Re\lambda_1)$ strictly larger than zero and $\lambda_1$ is the largest eigenvalue of the RG superoperator acting on local operators $\Phi(\mO_{UV})= V^\dagger U^\dagger \mO_{UV} U V $ \cite{kim2017entanglement}. 

\begin{figure}
    \centering
    \includegraphics[width=0.23\textwidth]{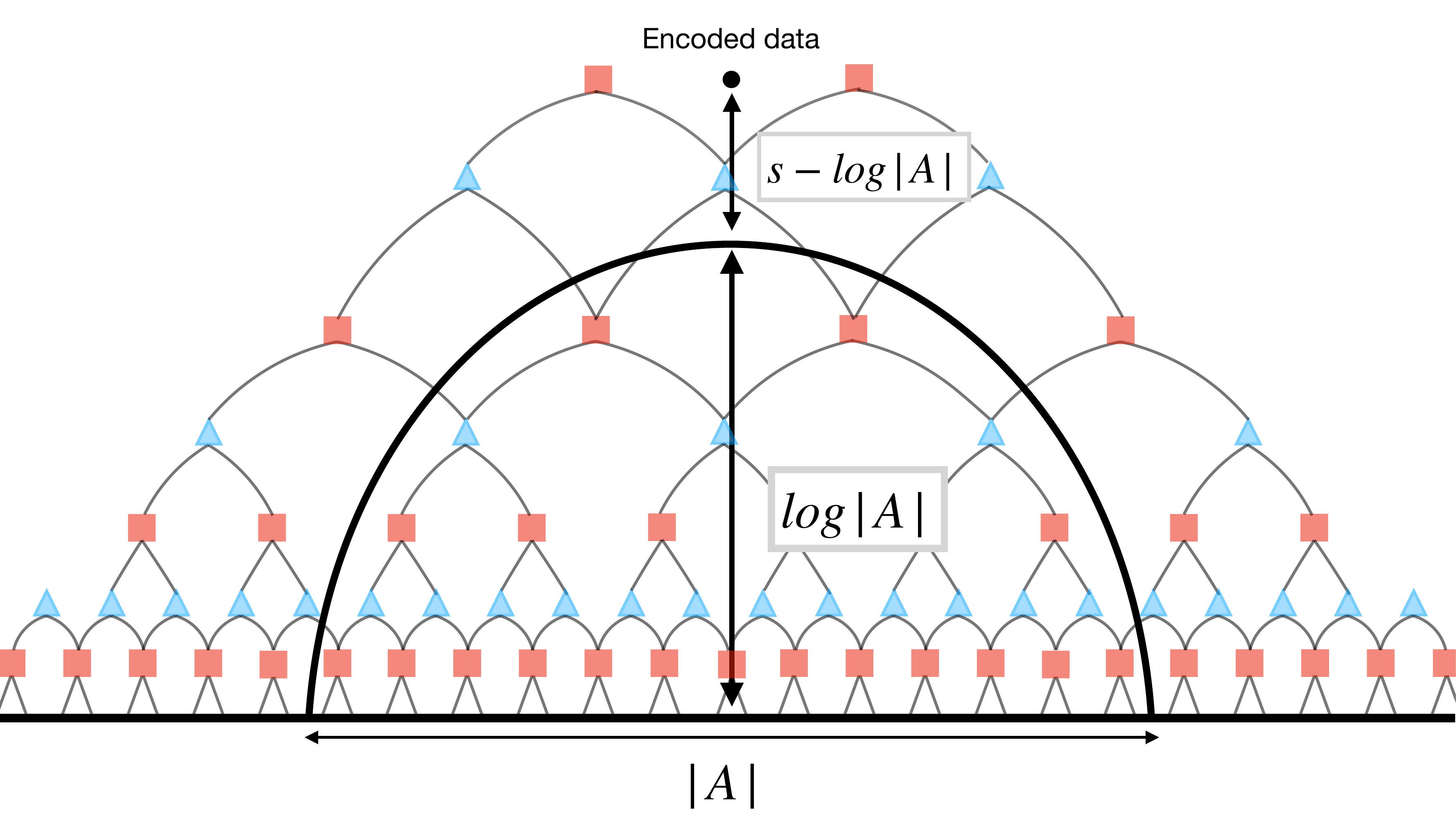}
    \includegraphics[width=0.23\textwidth]{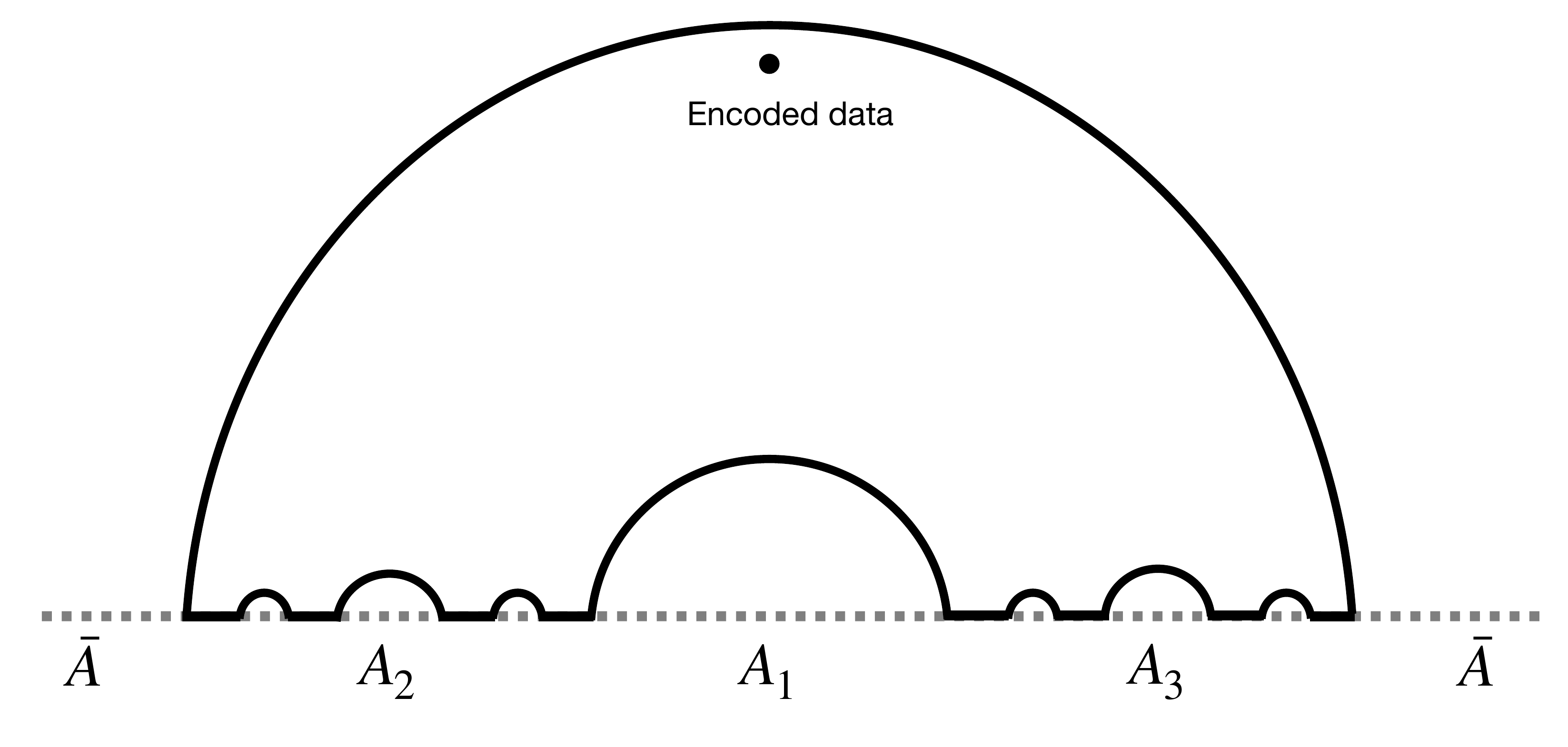}
    \caption{\small{(Left) The RG flow of operators localized in $A$ in MERA. The operator shrinks to a point after $s\sim\log|A|$. Deeper in the IR, its norm drops exponentially fast. (Right) The encoded data is protected against the erasure of multiple regions: $\bar{A}$, $A_1$, $A_2$, etc.}}
    \label{fig:mera_encode}
\end{figure}

In MERA, the local eigen-operators of the RG map are the conformal primaries and their corresponding eigenvalues fix the conformal dimensions. If each step of MERA cuts the number of sites down by a factor of $\gamma$ then
\begin{eqnarray}
\Phi(\mO_h)=\gamma^{-h}\mO_h
\end{eqnarray}
where $h$ is the conformal dimension of the primary operator $\mO_h$. As discussed in \cite{kim2017entanglement},
the RG evolution of non-local operators follows two stages: first the support of the operator shrinks to a single site, and then the norm of the local operator falls exponentially fast; see figure (\ref{fig:mera_encode} Left). The first stage takes approximately $\log |A|$ layers and after a total of $s$ layers in the second stage we have
\begin{eqnarray}
|\Phi^{s}(\mO)|\leq \gamma^{-\Delta (s-\log |A|)}|\mO|
\end{eqnarray}
where $\Delta$ is smallest conformal dimension which we refer to as the {\it gap} \footnote{It is called the gap because for a CFT on a sphere the dilatation is the Hamiltonian and $\Delta$ becomes the energy gap. In full generality, $\Delta$ need not be real, in which the inequality above should be replaced by $\Re(\Delta)$. Unitary tells us that $\Re(\Delta)\geq 0$. In a theory that flows in the IR to a conformal fixed point the expectation is that $\Delta$ is positive.}. Consider two code states $\ket{\Psi_r}$ and $\ket{\Psi_{r'}}$ in the infra-red after $s$ layers of RG. We have
\begin{eqnarray}\label{upperbound}
|\braket{\Psi_r|\Phi^s(\mO_i)|\Psi_{r'}}|\leq \ep_i=\gamma^{-\Delta(s-\log|A|)}|\mO_i|
\end{eqnarray}
where $\mO_i$ is any UV operator supported on region $A$.
For large $s$, the right-hand-side is small and this becomes the Knill-Laflamme condition for approximate error correction  \cite{brandao2019quantum}. For simplicity assume that every layer of MERA maps $N$ qubits to $N/2$ qubits; i.e $\gamma=2$. If we have $N$ qubits in the UV every qubit of the IR theory corresponds to an approximate error correction code that corrects the erasure of the simply connected region of size $d$ in the UV with errors controlled by the small parameter $\ep=2^{3-\Delta(s-\log |A|)/2}$.\footnote{In fact, we have universal subspace quantum error correction which means that any $2^k$ dimensional subspace of the IR theory is protected against the erasure of a region of size $|A|$ with errors $\ep=2^{1+2k-\Delta(s-\log |A|)/2}$.}

Given a code subspace the distance of a code $d$ is defined to be the minimum support of errors that cannot be corrected.
One might guess that the distance of the code we defined above is $|A|$. However, this is incorrect because in MERA there are smaller multi-component regions that contain the information content of the encoded qubit; see figure \ref{fig:mera_encode}. This is because we can erase $\bar{A}$ and smaller regions $A_1$, $A_2$ and $A_3$ inside $A$ and and still recover our information; see (Fig. (\ref{fig:mera_encode} Right). The authors of \cite{Pastawski:2016qrs} called this property uberholography.

% \begin{figure}
%     \centering
%     \includegraphics[width=0.6\linewidth]{PRL/uber.pdf}
%     \caption{\small{}}
%   \label{fig:uber}
% \end{figure}

\section{Continuous MERA}\label{sec:cMERA}

Continuous MERA(cMERA) is the generalization of MERA to continuum theories proposed in \cite{haegeman2013entanglement}. In the following, we go over the details of constructing cMERA of massive free boson field theory as in \cite{zou2019magic}.
%A generalization of MERA to continuum theories (cMERA) was proposed in \cite{haegeman2013entanglement}. 
% We first describe the cMERA in the Wilsonian picture with a sharp cut-off where the goal is to use the variational principle to fix the ground state at scale $\mu$ and cut-off $\Lambda$. 
Similar to the discrete case, cMERA is an isometric map that takes the states of a theory with zero correlation length deep in the IR and prepares the low energy states of a QFT (or CFT) in the UV. The IR ground state $\ket{\Omega^{(0)}}$ has no real-space entanglement. It is more convenient to think of MERA as the isometry from the IR to the UV. The state at energy scale $\Lambda e^{-s}$ is given by 
\begin{eqnarray}
\ket{\Omega^{(s)}}=\mathcal{P}e^{-i\int_{0}^s du (K(u)+L(u))}\ket{\Omega^{(0)}}
\end{eqnarray}
where $L(s)$ is the non-relativistic scaling transformation\footnote{It sends $x^i\to  \lambda x^i$ keeping time untouched.}, and $K(s)$ is the continuous analog of the layer of entanglers:
\begin{eqnarray}
K(s)=\int d^dk \:\Gamma(|k|/\Lambda) g(s,k)\mO_k\ .
\end{eqnarray}
Here, $\Lambda$ is the cut-off scale and the cut-off function $\Gamma(|k|/\Lambda)$ can be chosen to be sharp or smooth. The operator $\mO_k$ is an operator of energy scale $k$ that should be suitably chosen as the generator of the entangling layer. Finally, the function $g(s,k)$ decides the strength of the entangling procedure \cite{miyaji2015boundary}\footnote{In discrete MERA the effective cut-off changes as a function of scale, whereas in cMERA we have kept the cut-off $\Lambda$ fixed. To compare the two, we consider the rescaled state $\ket{\tilde{\Omega}(u)}=e^{iu L}\ket{\Omega(u)}$.} .
%For concreteness, we start with the example of free bosons.

For concreteness, consider free massive boson field $\phi(x)$ and its momentum conjugate $\pi(x)$ in one spatial dimension. Following \cite{zou2019magic} we choose an entangler independent of scale 
\begin{eqnarray}
K&=&\frac{\Lambda}{4}\int dxdy\: e^{-{\Lambda}|x-y|}\times\nn\\
&&(a(x;\Lambda)a(y;\Lambda)-a(x;\Lambda)^\dagger a(y;\Lambda)^\dagger)\nn
\end{eqnarray}
where 
\begin{eqnarray}
a(x;\Lambda)=\sqrt{\frac{\Lambda}{2}}\phi(x)+\frac{i}{\sqrt{2\Lambda}}\pi(x)
\end{eqnarray}
is the annihilation operator that defines the unentangled state via $a(x;\Lambda)\ket{\Omega^{(0)}}=0$. The cMERA state at scale $e^{-s}\Lambda$ is
\begin{eqnarray}
\ket{\Omega^{(s)}}=e^{is(L+K)}\ket{\Omega^{(0)}}\ .
\end{eqnarray}
This state is the ground state of a massive free boson deformed by a non-relativistic irrelevant term  \cite{zou2019magic}\footnote{Note that our convention differs from \cite{zou2019magic} in the sign of $s$.}:
\begin{eqnarray}
H^{(s)}(\Lambda)&=&\int \frac{dx}{2} \Big( \p_x\phi(x)^2+\pi(x)^2  \nn\\
&& +\Lambda^2 e^{2s}\phi(x)^2+\frac{1}{\Lambda^2}(\p_x\pi(x))^2 \Big) \nn \\
\end{eqnarray}
We define the annihilation operator that kills $\ket{\Omega^{(s)}}$: 
\begin{eqnarray}\label{massgeneral}
&&a_s(k;\Lambda)=\sqrt{\frac{\alpha_s(k;\Lambda)}{2}}\phi(k)+\frac{i}{\sqrt{2\alpha_s(k;\Lambda)}}\pi(k)\nn\\
&&\alpha_s(k;\Lambda)=\Lambda\sqrt{\frac{k^2+\Lambda^2e^{2s}}{k^2+\Lambda^2}}\ .
\end{eqnarray}
The mass term $m(s)=\Lambda e^{s}$ runs with scale vanishing in the UV $(s\to -\infty)$ and growing in the IR. 
% Under the RG the field operator is renormalized according to
% \begin{eqnarray}
% &&\phi(k;\Lambda,s_1)=\sqrt{\frac{\omega(k,s_1)}{\omega(k,s_2)}}\phi(k;\Lambda,s_2)\nn\\
% &&\pi(k,\Lambda,s_1)=\sqrt{\frac{\omega(k,s_2)}{\omega(k,s_1)}}\pi(k;\Lambda,s_2)
% \end{eqnarray}
% where $\omega(k,s)=\sqrt{k^2+m^2(s)}$. 
Since we are interested in the RG flow of a massive theory we fix the mass $m$ and vary the cut-off $\Lambda=m e^{-s}$. To further simplify our discussion we measure all dimensionful quantities in units of $m$ (we set $m=1$) \footnote{If instead of setting $m=1$ we take the massless limit in (\ref{massgeneral}) we find 
 \begin{eqnarray}
 \omega(k;\Lambda)=\frac{\Lambda}{\sqrt{k^2+\Lambda^2}}|k|\ .
 \end{eqnarray}
 Deep in the IR we have a CFT and the renormalized field operators $\phi^\Lambda(0)$, $\pi^\Lambda(0)$ and $V_p(0)=e^{ip\phi(0)}$ are conformal primaries satisfying
 \begin{eqnarray}
 -i[L+K,\mO^\Lambda_\alpha(0)]=\Delta_\alpha \mO^\Lambda_\alpha(0)
 \end{eqnarray}
 with conformal dimensions $\Delta_\phi=0$, $\Delta_\pi=1$ and $\Delta_{V_p}=p^2/2$. The field $\phi(x)$ is not really physical. Its vanishing conformal dimension is a symptom of the infra-red divergences in the two-point function of $\phi$.}:
%\paragraph*{Renormalization of field operators:}
%In magic cMERA, the annihilation operator of a massive field with cut-off $\Lambda$ is
\begin{eqnarray}\label{generala}
&&a_s(k)=\sqrt{\frac{\alpha_s(k)}{2}}\phi(k)+\frac{i}{\sqrt{2\alpha_s(k)}}\pi(k)\nn\\
&&\alpha_s(k)=\sqrt{\frac{k^2+1}{k^2e^{2s}+1}}\ .
\end{eqnarray}
% As we take the cut-off $e^{-s}$ to infinity we find the standard dispersion relation of massive fields, whereas in the limit $k\gg m$ and $k\gg \Lambda$ we find the annihilation operator corresponding to the unetangled state
% \begin{eqnarray}
% \psi(k)=\sqrt{\frac{\Lambda}{2}}\phi(k)+\frac{i}{\sqrt{2\Lambda}}\pi(k)\ .
% \end{eqnarray}
They satisfy the standard commutation relations \begin{eqnarray}
[a_s^\dagger(k),a_s(k')]=\delta_{kk'},
\end{eqnarray}
and the Hamiltonian is
\begin{eqnarray}
&&H^{(s)}=\int dk\:E_s(k)a_s^\dagger(k)a_s(k)\nn\\
&&E_s(k)=\sqrt{k^2+1}\sqrt{1+k^2 e^{2s}}\ .
\end{eqnarray}
% It is convenient to fix a cut-off scale $\Lambda_0=1$ and make the mass and momentum and the cut-off dimensionless $k\to k/\Lambda_0$ and $m\to m\to m/\Lambda_0$ and $\Lambda\to \Lambda_0 e^{-s}$. Then, under the RG flow of the massive theory towards the IR $s$ is increased from zero to infinity lowering the cut-off. Measured in units of $\Lambda_0$ the cut-off is at $e^{-s}$ and we have the redefined operator $a^{(s)}_m$ with the dispersion relation
% \begin{eqnarray}
% &&\omega^{(s)}_m(k):=\sqrt{\frac{k^2+m^2}{k^2e^{2s}+1}}\ .
% \end{eqnarray}
The renormalized creation/annihilation operators are related to those of the UV theory 
\begin{eqnarray}
&&a_s^\dagger(k)\pm a_s(k) =\beta_s(k)^{\pm 1}(a^\dagger(k)-a(k))\nn\\
%+\frac{(\beta(k,s)-\beta(k,s)^{-1})}{2}a^\dagger(k),\nn\\
%&&\phi^s(k)=\beta(k;s)^{-1}\phi(k),\qquad \pi^s(k)=\beta(k;s)\pi(k)\nn\\
&&\beta_s(k)=\lb 1+e^{2s} k^2\rb^{1/4}\ .
\end{eqnarray}
%Here, $e^s$ the cut-off length scale.
We denote by $\ket{\Omega^{(s)}}$ the vacuum state annihilated by the annihilation modes at scale $e^{-s}$.

%The coherent field operator is defined to be 
%\begin{eqnarray}
%&&D(f):=e^{a^\dagger(f)-a(f^*)},\qquad f=f_-+i f_+\nn\\
%&&a^\dagger(f_-+if_+)-a(f_--if_+)=(a^\dagger-a)(f_-)+i(a^\dagger+a)(f_+)\nn
%\end{eqnarray}
%where $f_{\pm}$ are real functions \footnote{For the construction of the coherent field operator see appendix \ref{app:coherent}.}. 
%The renormalization of the coherent operator can be absorbed in the choice of smoothing function $D(f)=D^{(s)}(f^s)
%$ with
%\begin{eqnarray}\label{rencoherent}
%&&f^s_\pm(k)=\beta_s(k)^{\pm 1}f_\pm(k),\nn\\
%&&f^s_\pm(x)=B^{\pm 1/4}f_\pm(x),\nn\\
%&&B:=(1-e^{2s}\p^2)\ .
%\qquad g^s_\pi(k)=\beta(k,s)^{-1}g_\pi(k)\nn\ .
%\end{eqnarray}
%Acting on the vacuum the coherent operator creates the coherent state with cut-off length $e^s$
%\begin{eqnarray}
%\ket{f;s}=D^{s}(f)\ket{\Omega^{(s)}}
%\end{eqnarray}
%which has the energy
%\begin{eqnarray}
%&&\braket{f;s|H^{(s)}|f;s}=\int dk\: E_s(k) |f(k)|^2\ .
% \nn\\
% &&=(C^{1/4}f|C^{1/4}f),\nn\\
% &&C=(\p^2+1)(1+e^{2s}\p^2)
%\end{eqnarray}
%It is convenient to define the following inner product on the space of test functions
%\begin{eqnarray}\label{testfunc}
%&&(f|g):=\int dx f^*(x) g(x)\ .
%\end{eqnarray}

\section{Coherent states in QFT}\label{sec:coherent}

Consider the canonical quantization of a free QFT in finite volume. The vacuum is the tensor product of the vacua corresponding to the annihilation operators 
\begin{eqnarray}
a(k)=\sqrt{\frac{\omega(k)}{2}}\phi(k)+ \frac{i\pi(k)}{\sqrt{2\omega(k)}}\ .
\end{eqnarray}
The coherent states are defined to be the eigenstates of the annihilation operator
\begin{eqnarray}\label{eigenannihilate}
a(k)\ket{f(k)}=f(k)\ket{f(k)}
\end{eqnarray}
where $f(k)=f_-(k)+i f_+(k)$ is a complex function of $k$. We define the functions $f_\phi$ and $f_\pi$ from the Fourier transform
\begin{eqnarray}
f_\phi(k)=\sqrt{2\omega(k)}f_+(k),\qquad f_\pi(k)=-\sqrt{\frac{2}{\omega(k)}}f_-(k)\nn\ .
\end{eqnarray}
The coherent states are prepared by acting on the vacuum with the displacement operator $D(\alpha):=e^{\alpha a^\dagger(k)-\alpha^* a(k)}$\footnote{An equivalent expression for the displacement operator is $D(\alpha)=e^{-|\alpha|^2/2}e^{\alpha a_k^\dagger}e^{-\alpha^* a_k}$. }
\begin{eqnarray}
\ket{f(k)}&=&D(f(k))\ket{\Omega_k}\nn\\
&=&e^{i(f_\phi(k)\phi(k)+f_\pi(k) \pi(k))}\ket{\Omega_k}\ .
\end{eqnarray}
The displacement operators satisfy the Weyl algebra 
\begin{eqnarray}
D(\alpha)D(\beta)=e^{\frac{1}{2}(\alpha\beta^*-\alpha^*\beta)}D(\alpha+\beta)\ .
\end{eqnarray}
The coherent states form an overcomplete basis with the overlaps given by
\begin{eqnarray}\label{overlap}
\braket{g(k)|f(k)}=e^{-\frac{1}{2}(|f(k)|^2+|g(k)|^2-2g(k)^*f(k))}\ .
\end{eqnarray}

Taking the tensor product over all momentum modes our multi-mode coherent state is parametrized by an arbitrary complex function $f(x)$:
\begin{eqnarray}
&&\ket{f}=D(f)\ket{\Omega},\qquad f=f_-+i f_+\nn\\
&&D(f)=\otimes_k D(f(k))=e^{a^\dagger(f)-a(f^*)} \nn\\
&&=e^{i\phi(f_\phi)+i\pi(f_\pi)}=e^{i\phi(f_\pi)}e^{i\pi(f_\pi)}e^{i(f_\phi|f_\pi)} \
\end{eqnarray}
where we have used the Baker-Hausdorff-Campbell for $[X,Y]\sim \mathbb{I}$: $e^{X+Y}=e^Xe^Ye^{-\frac{1}{2}[X,Y]}$.
It follows from (\ref{overlap}) that the expectation value of the multi-mode displacement operator in the vacuum is
\begin{eqnarray}
&&\braket{D(f)}=\braket{\Omega|f}=e^{-\frac{1}{2}(f|f)}\nn\\
&&(f|g)\equiv\int dx f(x)^*g(x)\ .
\end{eqnarray}
% The Hilbert space overlaps of coherent states are given by 
% \begin{eqnarray}
% \braket{h|g}&=&\braket{\Omega|e^{i\phi(h-g)}|\Omega}\nn\\
% &=&e^{-\frac{1}{2}(h-g|h-g)_\phi}
% \end{eqnarray}
% where we have defined the following inner product on the space of test functions:
% \begin{eqnarray}
% &&(h|g)_\phi=\sum_k\tilde{f}_k\tilde{h}_k,\qquad \tilde{f}_k\equiv\frac{f_k}{\sqrt{2\omega_k}}
% \end{eqnarray}
These operators satisfy the Weyl algebra
\begin{eqnarray}
&&D(f)D(g)=e^{\frac{1}{2}\lb(g|f)-(f|g)\rb}D(f+g)\nn\\
&&=e^{i\lb (g_-|f_+)-(g_+|f_-)\rb}D(f+g)=e^{2i\text{Im}(g|f)}D(f+g)\nn
\end{eqnarray}
where we have separated the real and imaginary part of $f$ and $g$: $f=f_-+i f_+$ and $g=g_-+ig_+$.
It follows from (\ref{eigenannihilate}) that
% \begin{eqnarray}
% \lb\phi_k+\frac{i\pi_k}{2\omega_k}\rb\ket{h}=\frac{i h_k}{2\omega_k}\ket{h}\ .
% \end{eqnarray}
% As a result, we find
\begin{eqnarray}
e^{a(h^*)}\ket{f}=e^{(h|f)}\ket{f}\ .
\end{eqnarray}
If the Hamiltonian is
\begin{eqnarray}
H=\int dk\:E(k) a^\dagger(k)a(k)
\end{eqnarray}
then the energy of the field coherent state is
\begin{eqnarray}
\braket{f|H|f}=\int dk \: E(k)|f(k)|^2\ .
\end{eqnarray}

\section{The RG flow of the coherent operator}\label{sec:RGcoherent}

The vacuum state of the massive QFT with the cut-off length scale $e^{s}$ satisfies
\begin{eqnarray}
a_s(k)\ket{\Omega^{(s)}}=0
\end{eqnarray}
for the annihilation operators at scale $e^s$ defined in (\ref{generala}). 
The coherent state corresponding to this annihilation operator is
\begin{eqnarray}
&&\ket{f;s}=D^{(s)}(f)\ket{\Omega^{(s)}}\nn\\
&&D^{(s)}(f)=e^{a_s^\dagger(f) -a_s(f^*)}
% \nn\\
% &&=e^{i \phi(\tilde{f}^\Lambda_s)+i \pi(\tilde{f}^\Lambda_s)}=D(f^\Lambda_s)
\end{eqnarray}
It satisfies the Weyl algebra
\begin{eqnarray}\label{Weylalgebra}
D^{(s)}(f)D^{(s)}(g)=e^{2i\text{Im}(f|g)}D^{(s)}(f+g)\ .
\end{eqnarray}
similar to the set of coherent operators $D(f)$ in the UV.
% This multiplication rule can be used to compute the three
% %  We define the functions $f^{(s)}=f^{(s)}_\pi+if_{\phi}^{(s)}$
%  % \begin{eqnarray}
% % % % &&\tilde{f}_{s,\phi}^\Lambda(k)=\sqrt{2\omega^\Lambda_s(k)}f_\phi(k),\qquad \tilde{g}_{s,\pi}^\Lambda(k)=-\sqrt{\frac{2}{\omega^\Lambda_s(k)}}g_\pi(k)\nn\\
% % % &&f^\Lambda_{s,\phi}(k)=\beta_s(k,\Lambda)f_\phi(k),\qquad g^\Lambda_{s,\pi}(k)=\beta_s(k,\Lambda)^{-1}g_\pi(k)\nn\\
% % % &&\beta_s(k,\Lambda)=\sqrt{\frac{\omega^\Lambda_{s_2}(k)}{\omega^\Lambda_{s_1}(k)}}
% % % \end{eqnarray}
%  such that
%  \begin{eqnarray}
%  D(f)=D^{(s)}(f^{(s)})\ .
% \end{eqnarray}
% If we multiply two coherent operators corresponding to different scales we find
% \begin{eqnarray}
% D^\Lambda_{s_1}(f)D^\Lambda_{s_2}(g)=e^{2i\Im(g^\Lambda_{s}|f)}D^\Lambda_{s_1}(f+g ^\Lambda_{s})\nn\ .
% \end{eqnarray}
The matrix element of $D(g)$ in the code states correspond to a three-point function of coherent operators that can be computed using the multiplication rule of the algebra in (\ref{Weylalgebra}): 
\begin{eqnarray}\label{threept}
e^{A(p',p;s)}&:=&\braket{p'f,s|D(g)|pf,s}\nn\\
&&=\braket{\Omega^{(s)}|D^{(s)}(-p'f)D^{(s)}(g^s)D^{(s)}(pf)|\Omega^{(s)}}\nn\\
&&=e^{2i\text{Im}\lb -(g^s|p'f)+(pf|g^s-p'f)\rb}\times\nn\\
&&\braket{\Omega^{(s)}|D(g^s+(p-p')f)|\Omega^{(s)}}\ .
\end{eqnarray}
%where $\delta s=s_2-s_1$.
In the case where $f=if_0$ and $g=g_-+i g_+$:
 \begin{eqnarray}
 A(p',p;s)&=&-i(p+p')(g^s_-|f_0)\nn\\
 &&-\frac{1}{2}(g^s+(p-p')f_0|g^s+(p-p')f_0)\nn\ .
\end{eqnarray}

It is quite useful to know that the renormalization of the coherent operator can be absorbed in the choice of smooth function $D(f)=D^{(s)}(f^s)$ with
\begin{eqnarray}\label{rencoherent}
&&f^s_\pm(k)=\beta_s(k)^{\pm 1}f_\pm(k),\nn\\
&&f^s_\pm(x)=B^{\pm 1/4}f_\pm(x),\nn\\
&&B:=(1-e^{2s}\p^2)\ .
%\qquad g^s_\pi(k)=\beta(k,s)^{-1}g_\pi(k)\nn\ .
\end{eqnarray}

\subsection{Renormalization of coherent operators:}

We would like to understand the renormalization of the coherent operators as they flow from the UV to the IR. 
As an example, consider the test function $f_0(x)$ in (\ref{Gaussian}) and the coherent operator $D(f_{0}+if_{0})$. Under the RG flow it goes to $D^{(s)}(f^s_{0,-}+if^s_{0,+})$ with $f^s_{0,\pm}$ real. It follows from (\ref{rencoherent}) that
 \begin{eqnarray}\label{renf0}
 &&f^s_{0,\pm}(x)=(1-e^{2s}\p_x^2)^{\pm 1/4} \frac{e^{-\frac{(x-x_0)^2}{2\ep^2}}}{\sqrt{2\pi}\ep}\ .
  \end{eqnarray}
 Deep in the UV the term $e^{2s}\p_x^2$ is small and the renormalization of $f_0$ is perturbative. The renormalization becomes non-perturbative at the cut-off length scale $e^s$ when $e^{2s}|\p_x^2 f_{0}|$ becomes comparable to $|f_{0}|$. For the test function $f_0$ we have
  \begin{eqnarray}\label{derivs}
  \frac{e^{2s}|\p_x^2 f_{0}|}{|f_{0}|}=\frac{e^{2s}}{\ep^2}\lb \frac{(x-x_0)^2}{\ep^2}-1\rb\ .
  \end{eqnarray}
%This shows that for the points near the peak $|x-x_0|\ll \ep$ in $f^s_{0}$ the term $(e^{2s}\p_x^2)$ in (\ref{renf0}) dominates.

%grow (decays) exponentially in $s$, respectively. 
%   Since the functions $f^s_{0,\pm}(k)$ are positive-definite it follows from the Bochner's theorem that its Fourier transform is a positive measure:
%   \begin{eqnarray}
%   f_{0,\pm}^s(x)\geq 0\ .
%   \end{eqnarray}
%  The zero mode of $f^s_{0,\pm}(s)$

\begin{figure}
    \centering
    \includegraphics[width=0.23\textwidth]{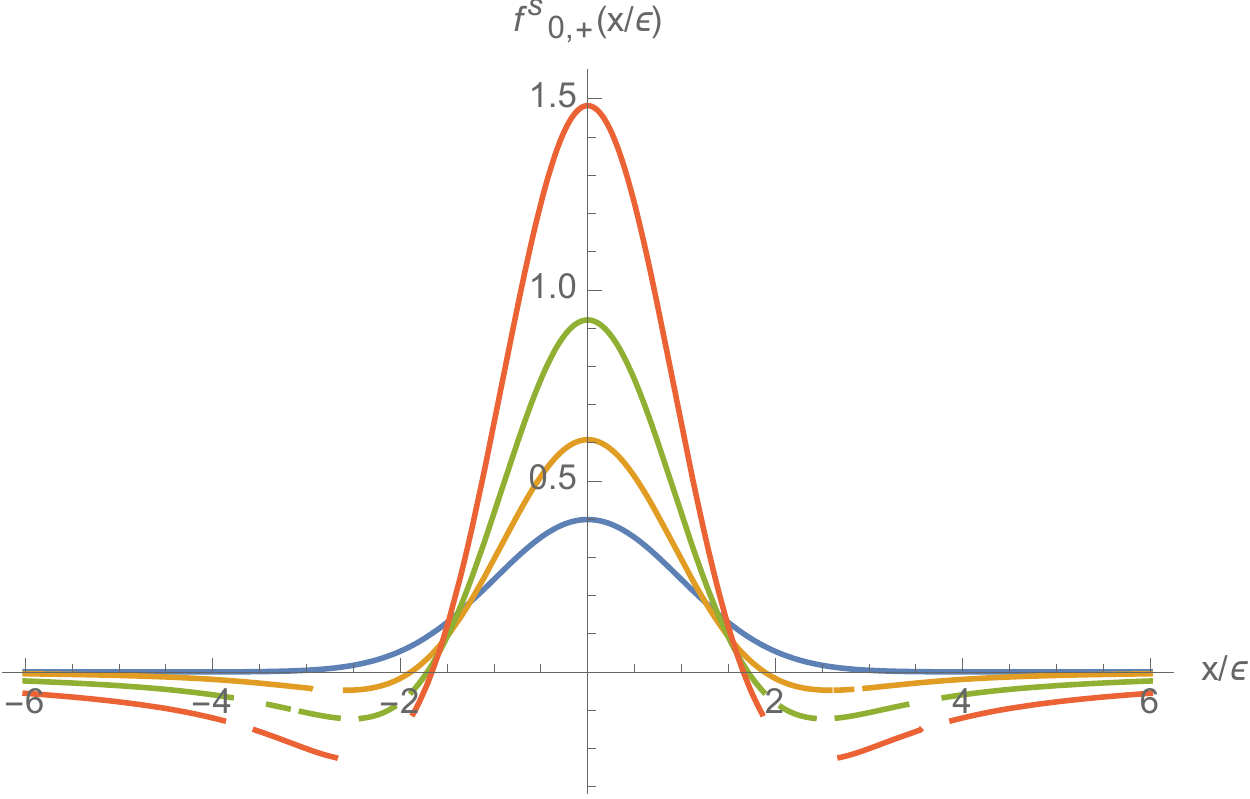}
    \includegraphics[width=0.23\textwidth]{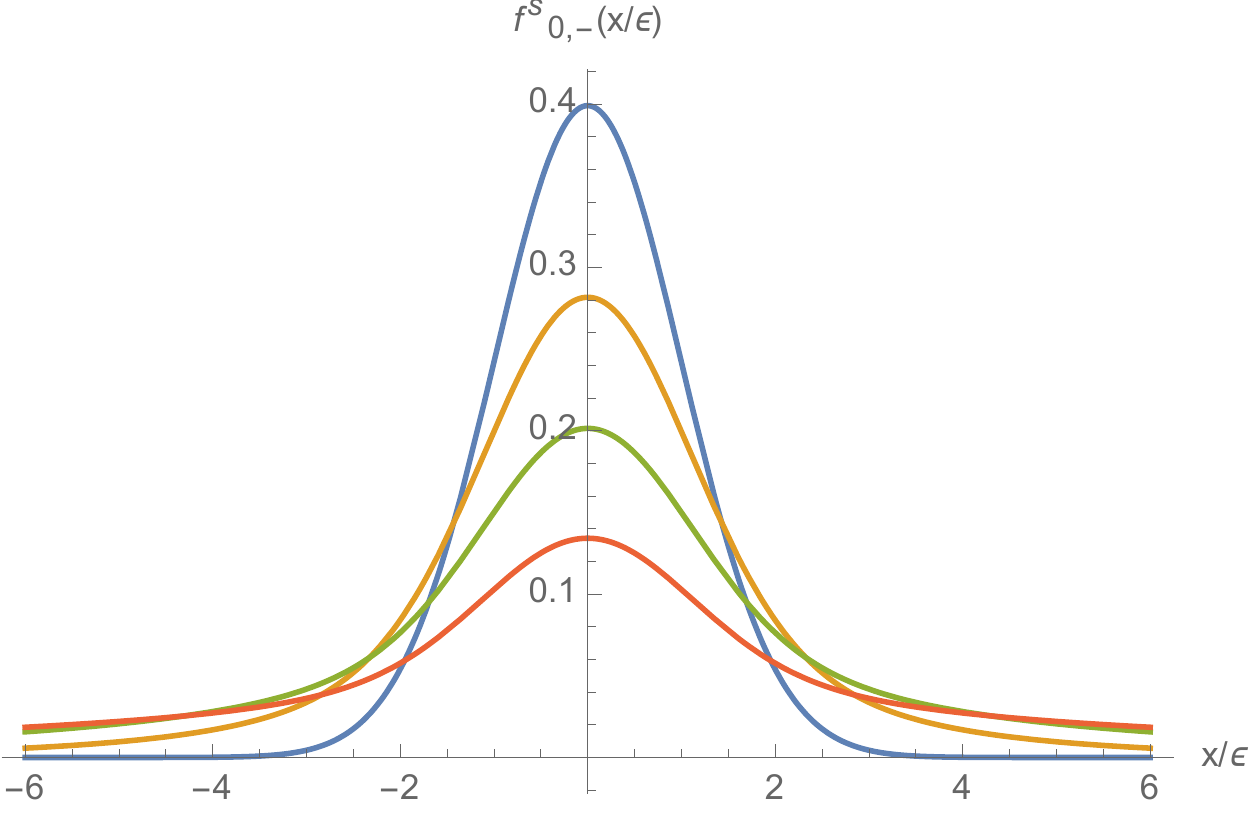}
    \caption{\small{The renormalization of the functions $f^s_{0,\pm}(x)$ is insignificant until the cut-off length scale becomes comparable to $\ep$ (width of $f_0(x)$). As we flow further towards the IR (Left) the function $f^s_{0,+}(x)$ becomes highly peaked (Right) the function $f^s_{0,-}(x)$ flattens out. (Blue: $s=-\infty$, Yellow: $s=1$, Green: $s=2$, Red: $s=3$.)}}
    \label{fig:numerical}
\end{figure}

  There are two stages to the RG flow of this coherent operator.
  In the first stage, the cut-off length $e^s$ is much smaller than $\ep$, the term $e^{2s}\p_x^2$ in (\ref{renf0}) can be neglected and the renormalization of $f^s_{0}$ is perturbatively small. The second stage starts when $e^s\sim \ep$. As we flow deeper in the IR $e^{s}\gg \ep$ the  term $e^{2s}\p_x^2$ in (\ref{renf0}) dominates. 
  In stage two, we are in the regime $\ep\ll e^{s}$ and for points away from $|x-x_0|=\ep$ we can use the approximation
%   \footnote{For a more rigorous justification of this approximation see appendix \ref{}\NL{AddAppendix}.}
  \begin{eqnarray}\label{approx}
  &&f^s_{0,\pm}(x)\simeq e^{\pm s/2}(\p_x^2)^{\pm 1/4}f_0(x)\ .
%   \nn\\
%   &&f^s_{0,+}(x)=\frac{e^{s/2}e^{-\tilde{x}^2}}{\sqrt{2\ep^3|\tilde{x}|}}\lb \tilde{x}^2I_{5/4}(\tilde{x}^2)-(\tilde{x}^2-1/2)I_{1/4}(\tilde{x}^2)\rb\nn\\
%   &&f^s_{0,-}(x)=\frac{e^{-s/2}e^{-\tilde{x}^2}\sqrt{|\tilde{x}|}}{\sqrt{2\ep}|\tilde{x}|}I_{-1/4}(\tilde{x}^2)
  \end{eqnarray}
%   where $\tilde{x}=x/\ep$. 
  The function $f^s_{0,+}$ ($f^s_{0,-})$ grows (decays) exponentially fast as $e^{(s-\log\ep)/2}$ $(e^{-(s-\log\ep)/2})$ in the IR, respectively; see (Fig. \ref{fig:numerical}). 
  
We can generalize these lessons to the renormalization of any function $g(x)$ that is localized around $x=x_0$ with linear size $|A|$. Intuitively, one expects that near the peak
\begin{eqnarray}\label{generalfunc}
\frac{|\p_x^2g(x)|}{|g(x)|}\Big|_{x\simeq x_0}=O(|A|^{-2})
\end{eqnarray}
or more generally the right-hand-side is some function that is inversely proportional to $|A|$. There are two stages to the RG flow. In the first stage $e^s\ll |A|$, the cut-off grows but the function is frozen. In comparison to MERA, the cut-off can be interpreted as a unit qudit and the operator is supported on $|A|/\ep$ number of sites. Therefore, in this stage the support of the UV operator shrinks exponentially fast. Similar to MERA, the second stage starts when the RG scale reaches the size of the unit block $e^s\simeq |A|$. Beyond this point, we find that inside $A$ the function $g^s_+$ grows exponentially as $e^{(s-\log|A|)/2}$ and $g^s_-$ decays exponentially as $e^{-(s-\log|A|)/2}$.  This is reminiscent of the second of the RG flow of operator in MERA. For a general function $g(x)$ the right-hand-side of (\ref{generalfunc}) is some more complicated function. The transition scale happens at some $e^s\sim h(|A|)$ for some increasing positive function of $|A|$ and the exponent that controls the exponential growth or decay is $s-\log h(|A|)$.

\section{Quantum error-correction in $c$MERA}

In this section, we provide the explicit calculation of encoding a qudit system into a massive free boson field theory.

\subsection{Encoding a qudit at a point:}\label{encoding}

Consider the Gaussian wave-packet
\begin{eqnarray}\label{Gaussian}
f_0(x)=\frac{1}{\sqrt{2\pi}\ep}e^{-\frac{(x-x_0)^2}{2\ep^2}}
\end{eqnarray}
that is a regularization of the Dirac delta function \footnote{In the momentum space we have $f_0(k)=\frac{e^{i k x_0-\frac{\ep^2 k^2}{2}}}{\sqrt{2\pi}}$.}:
 \begin{eqnarray}
  &&\lim_{\ep\to 0}(g|f_0)=g(x=x_0),\qquad (f_0|f_0)=\frac{1}{2\sqrt{\pi}\ep}\ .
 \end{eqnarray}
%  in the infinite volume limit (continuous $k$ variable).\footnote{Note that Fourier transform of this function to real-space is real: $f_a(x)=\pi^{-1/4}\ep^{-1/2} e^{(x-a)^2/(2\ep^2)}$.} 
%  As we make $\ep$ smaller the wave-packet gets highly localized on $x=a$; see figure \ref{wavepack}.
%  \begin{figure}
%     \centering
%     \includegraphics[width=.8\linewidth]{PRL/wavepacket.pdf}
%     \caption{The functions $f_a(x)$ becomes highly peaked around $x=a$ as we send $\ep\to 0$.}
%     \label{wavepack}
% \end{figure}
 We consider two types of unitary coherent operators $D(ipf_0)$ and $D(g_-)$ for $p$ integer and $g_-$ a real function. We define the code states
 \begin{eqnarray}
 \ket{p,x_0}:=D(ipf_0)\ket{\Omega}\ .
 \end{eqnarray}
 These states have large energy at small $\ep$:
 \begin{eqnarray}
 \braket{p,x_0|H|p,x_0}=\frac{p^2}{2\ep^4\sqrt{\pi}}U(-\frac{1}{2},0,\ep^2)\simeq \frac{p^2}{2\pi\ep^4}
 \end{eqnarray}
 where $U(a,b,z)$ is the confluent hypergeometric function.
 The set of states $\ket{p,x_0}$ are almost orthonormal in the limit of small $\ep$ because  
\begin{eqnarray}
\braket{p',x_0|p,x_0}&=&\braket{D(i(p-p')f_0)}=e^{-\frac{1}{2}(p-p')^2(f_0|f_0)}\simeq\delta_{pp'}\nn
\end{eqnarray}
% where  
% \begin{eqnarray}
% (f|g)=\int dx\: f^*(x)g(x)\ .
% \end{eqnarray}
If $p$ runs from $0$ to $q-1$ the almost orthonormal states $\ket{p,x_0}$ encode a $q$-level quantum system at $x=x_0$.
The operator $D(iqf_0)$ takes us in between code states
\begin{eqnarray}\label{logicalmom}
\braket{p',x_0|D(i qf_0)|p,x_0}=\delta_{p',p+q}
\end{eqnarray}
and the operator $D(g_-)$ is diagonal in the basis $\ket{p,x_0}$:
\begin{eqnarray}
\braket{p',x_0|D(g_-)|p,x_0}&=&\delta_{pp'}\braket{D(g_-^s)}e^{-i(p+p')(g_-|f_0)}\nn\ .
%&=&\delta_{pp'}e^{-i(p+p')q}\ .
\end{eqnarray}
where $(g_-|f_0)\simeq g_-(x=x_0)$ in the $\ep\to 0$ limit. If $P_0$ is the projection to the code subspace spanned by $\ket{p,x_0}$ then the operators $P_0$, $P_0 D(iqf_0)P_0$ and $P_0D(g_-)P_0$ and their Hermitian conjugates generate the algebra of the $q$-level system encoded at point $x=x_0$.

Moving a distance $\ep$ away from $x=x_0$ we can encode a new $q$-level system because
\begin{eqnarray}
\braket{p',x_0|p,x_1}=&&\braket{D(ipf_1-ip'f_0)}\nn\\
=\exp&&\lb -\frac{1}{2}(f_0|f_0)\lb p^2+(p')^2-2pp'e^{\frac{-(x_0-x_1)^2}{4\ep^2}}\rb\rb\nn
\end{eqnarray}
% where we have used 
% \begin{eqnarray}
% (f_0|f_b)&=&(f_a|f_a) e^{-\frac{(a-b)^2}{4\ep^2}}\nn\ .
% \end{eqnarray}
which is vanishing small at small $\ep$ and $|x_0-x_1|>\ep$.

 \subsection{Error correction condition} 
 
%Consider the Gaussian wave-packet
%\begin{eqnarray}\label{eq:Gaussian}
%f_0(x)=\frac{1}{\sqrt{2\pi}\ep}e^{-\frac{(x-x_0)^2}{2\ep^2}}
%\end{eqnarray}
%as the smooth function for our encoding.
% \footnote{This is a regularization of the Dirac delta function:
% % In the momentum space we have $f_0(k)=\frac{e^{i k x_0-\frac{\ep^2 k^2}{2}}}{\sqrt{2\pi}}$. 
% \begin{eqnarray}
% &&\lim_{\ep\to 0}(g|f_0)=g(x=x_0),\qquad (f_0|f_0)=\frac{1}{2\sqrt{\pi}\ep}\ .
% \end{eqnarray}}.\KF{This is a regularization of the Dirac delta function:
% % In the momentum space we have $f_0(k)=\frac{e^{i k x_0-\frac{\ep^2 k^2}{2}}}{\sqrt{2\pi}}$. 
% \begin{eqnarray}
% &&\lim_{\ep\to 0}(g|f_0)=g(x=x_0),\qquad (f_0|f_0)=\frac{1}{2\sqrt{\pi}\ep}\ .
% \end{eqnarray}}
Below, we use the algebra of the free fields to compute these matrix elements;
\begin{eqnarray}\label{exponent}
&&\braket{p',x_0;s|D(h_-)|p,x_0;s}\simeq c(h_-,s)\delta_{pp'}e^{-i(p-p')(h^s_-|f_0)}\nn\\
&&\braket{p',x_0;s|D(ih_+)|p,x_0;s}=e^{-\frac{1}{2}(h^s_++(p-p')f_0|h^s_++(p-p')f_0)}\nn\\
&&c(h_\pm,s)=\braket{\Omega^s|D(h_\pm)|\Omega^s}\ .
% nn\\
% &&c(f_0,s)=\braket{\Omega^s|D(if_0)|\Omega^s}\ .
\end{eqnarray}
%and we have used $e^{-\frac{(p-p')^2}{4\pi\ep}}\simeq \delta_{pp'}$ for small $\ep$.
Here, we use $h_-$ and $h_+$ instead of $g_-$ and $qf_0$ which could be supported on either A, C, or both to distinguish the smooth functions for the error operators from the ones used for the encoding $f_0$ .

%\begin{figure}
 %   \centering
  %  \includegraphics[width=0.6\linewidth]{PRL/gc.pdf}
   % \caption{\small{}}
   %\label{fig:gc}
%\end{figure}

\begin{figure}
   \centering
   \includegraphics[width=1.0\linewidth]{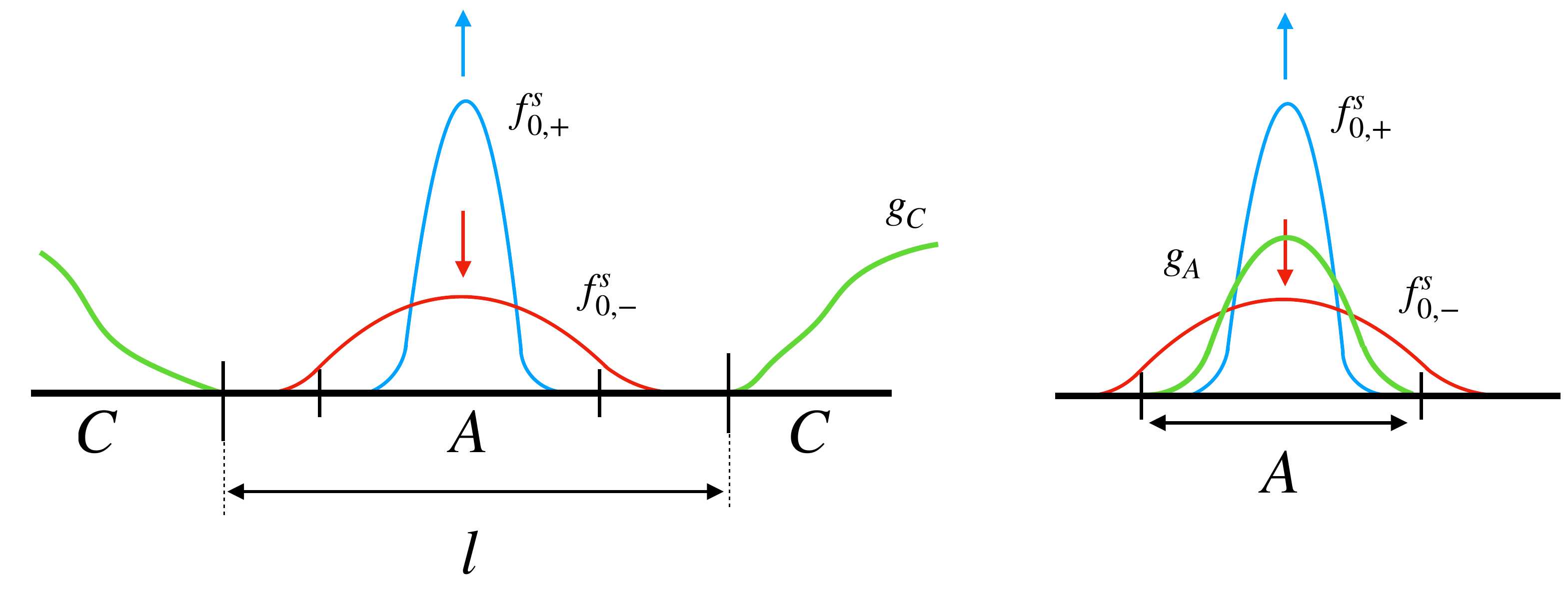}
   \caption{\small{Under the RG $f^s_{0,+}$ $(f^s_{0,-})$ localizes (flattens). (Left) Their overlap with $g_C$ supported on $C$ is suppressed if $l$ is large enough. (Right) The overlap of $f^s_{0,+}$ $(f^s_{0,-})$ with $g_A$ supported on $A$ grows (decays), respectively.}}
   \label{fig:gagc}
\end{figure}

First, consider the case where $h_C=h_{-,C}+i h_{+,C}$ with $h_{\pm,C}$ real functions  supported on a region $C$ that is at least $l\ep$ away from $x=x_0$; see (Fig. \ref{fig:gagc} Left).
We have 
\begin{eqnarray}\label{invert}
&&(h^s_{\pm}|f_{0})=(h_{\pm}|f^s_{0,\pm})\ . 
%&&(g^s_{+,A}|f_0)=O(e^{-\frac{l^2}{2}})\ .
\end{eqnarray}
Since $f^s_{0,-}(x)$ decays exponentially fast with distance away from $x=x_0$ we have
\begin{eqnarray}
(h^s_{\pm, C}|f_0)=O(e^{-l^2/2})\ .
\end{eqnarray}
Since all the points in $C$ satisfy $|x-x_0|\gg \ep$ it follows that the functions $f_{0,\pm}(x)$ grow at most like $e^{s/2}/\sqrt{\ep}$. 
% \begin{eqnarray}
% &&(g^s_{-,C}|f_0)=O(e^{-\frac{1}{2}(l^2-(s-\log\ep))})\nn\\
% &&(g^s_{+,C}|f_0)=O(e^{-\frac{l^2}{2}})\ .
% \end{eqnarray}
For any fixed large $s$ and a large enough $l$ we find that the error caused by any $D(h_C)$ can be corrected because
\begin{eqnarray}
&&\braket{p',x_0;s|D(h_{-,C})|p,x_0;s}\simeq c(h_{-,C},s)\delta_{pp'}\nn\\
&&\braket{p',x_0;s|D(ih_{+,C})|p,x_0;s}\simeq c(h_{+,C},s)\delta_{pp'}\ .
\end{eqnarray}

Next, we consider the case where $h$ is supported on region $A$  that includes the point $x=x_0$; see (Fig. \ref{fig:gagc}). As we argued, under the RG flow, the function $h^s_A$ is unchanged until $e^s\sim |A|$. After that the second stage of RG starts.
% we use the fact that 
% \begin{eqnarray}
% &&(g^s_{\pm,A}|f_0)=(g_{\pm,A}|f^s_{0,\pm})\ . 
% %&&(g^s_{+,A}|f_0)=O(e^{-\frac{l^2}{2}})\ .
% \end{eqnarray}
Since $e^{s}\gg |A|$, for all $(x-x_0)\ll |A|$, we can use the approximation in (\ref{approx}). Therefore, in the second stage we have
\begin{eqnarray}
(h_{\pm, A}|f^s_{0,\pm})\simeq e^{\pm 1/2(s-\log|A|)}(h_{\pm, A}|(\p_x^2)^{\pm 1/4}f_0)\ .
\end{eqnarray}
This combined with (\ref{invert}) and (\ref{exponent}) implies that for $e^s\gg |A|$ we have
\begin{eqnarray}
&&\braket{p',x_0;s|D(h_{-,A})|p,x_0;s}\simeq c(h_{-,A},s)\delta_{pp'}e^{-i(p-p')O(e^{-s/2})}\nn\\
&&\simeq c(h_{-,A},s)\delta_{pp'}\ .
\end{eqnarray}
For the operator $D(ih_{+,A})$ as we saw in equation (\ref{logicalmom}) we can distinguish different code states if we tune $h_{+,A}=-(p-p')f_0$ so that the exponent in the second line of (\ref{exponent}) vanishes. However, this cancellation does not survive under the RG flow because
\begin{eqnarray}
&&h^s_{+,A}+(p-p')f_0=(p-p')(f_0-f^s_{0,+})\nn\\
&&=(p-p')(1-(e^{2s}\p_x^2)^{1/4})f_0 \simeq (p'-p)(e^{2s}\p_x^2)^{1/4}f_0\nn
\end{eqnarray}
where in the last line we use the fact that we are in a regime where $(e^{2s}\p_x^2)^{1/4}$ dominates over the first term. It is clear from the second line of (\ref{exponent}) that the norm of the function above controls the size of the matrix element of this coherent operator. This norm can be computed explicitly:
\begin{eqnarray}
(p-p')^2e^s(\p_x^{1/2}f_0|\p_x^{1/2}f_0)=(p-p')^2 \frac{e^{s}}{2\pi \ep^4}
\end{eqnarray}
Since $e^s\gg |A|$ and $|A|\gg 1$ the expression above grows to infinity. Plugging this back into (\ref{exponent}) we find that deep in the IR
\begin{eqnarray}
&&\braket{p',x_0;s|D(i(p-p')f_0)|p,x_0;s}=e^{-\frac{(p-p')^2}{2}O(e^s)}
\end{eqnarray}
which goes to zero.

\section{Holographic RG flows and phase transition}\label{sec:holography}

In this section, first, we review the basics of holographic QEC and the definitions of price, distance introduced in \cite{Pastawski:2016qrs}. Then, we discuss the two inequalities, no free lunch and holographic Singleton bound. In the next section, we explain the cause of the violation in the context of holographic RG followed by the subsection providing the detailed calculations of the critical interval size where the 'phase transition' happens. The last section reviews a reconstruction wedge and restate the conclusion in the main text for completeness.

\subsection{Holographic singleton bound}

The AdS-CFT correspondence is a holographic duality according to which the information about the states of a theory of gravity in an asymptotically locally AdS spacetime is encoded on the state of a CFT living on the boundary of the spacetime. It was argued in \cite{Almheiri:2014lwa} that the way holographic encoding works is reminiscent of operator algebra quantum error correcting code. In particular, the \textit{bulk} (gravitational) operators should be thought of as the logical operators which are mapped to physical \textit{boundary} (CFT) operators through the holographic duality. Since then, various realizations of the holographic error correcting codes have been proposed using the tensor networks \cite{Pastawski:2015qua,Yang:2015uoa,Hayden:2016cfa}.

%A bulk operator can be represented as a non-local boundary operator. The main insight from [CITE] was that we xxx. 

A refined version of the correspondence, called the `subregion-subregion duality', is a correspondence between a subregion $B$ of the boundary and a corresponding bulk region $\mathcal{E}(B)$. It asserts that the logical subalgebra corresponding to some bulk subregion can be \textit{reconstructed} from the subalgebra of the boundary subregion $B$. The bulk region $\mathcal{E}(B)$ is the domain of dependence of a bulk codimension-$1$ spacelike surface between $B$ and a bulk codimension-$2$ stationary area surface anchored on $B$ \cite{Dong:2016eik}. In case there are more than one possible stationary area surface anchored on $B$, the surface with the smallest area will determine the bulk region $\mathcal{E}(B)$. The bulk region $\mathcal{E}(B)$ is called the entanglement wedge of $B$ \cite{Headrick:2014cta} whereas the minimal stationary area surface is called Hubeny-Ryu-Takayanagi surface (HRT) surface \cite{Ryu:2006bv,Hubeny:2007xt}.

%for a subregion $B$ of the boundary, there is a corresponding subregion $b$ of the bulk spacetime such that a logical bulk operator in the $b$ can be represented as a non local operator in $B$. The bulk region $b$ is the entanglement wedge of the boundary region $B$. %The bulk region $b$ is called the entanglement wedge of $B$. and is defined with respect to the Ryu-Takayanagi (RT) surface of a boundary region $B$, which is a codimension-$2$ minimal\footnote{In a generic time dependent bulk geometry, the appropriate surface is a stationary area surface instead of a minimal area surface \cite{HRT}. However, we are only going to consider time-independent geometry in this work and hence, will only focus on minimal area surfaces.} area on the bulk spacetime anchored on the boundary region $B$. If there are many minimal area surface anchored to $B$, the RT surface is the one with the minimum area. The bulk subregion $b$ for a boundary subregion $B$ is the region bounded by the $B$ and RT surface corresponding to $B$\footnote{Technically speaking, the region $b$ is the bulk domain of dependence of a bulk region bounded by $B$ and the RT surface corresponding to $B$. However, we will only look at a fixed time slice of the bulk and hence ignore this distinction.}.  

Various properties of the holographic QECC were studied in \cite{Pastawski:2016qrs} and the authors defined the concept of the distance and the price for a logical algebra of operators in a bulk region. Here, we review some important definitions and theorems from \cite{Pastawski:2016qrs}. 

We start by considering a logical algebra of a single bulk point $x$. The distance of a logical operator is the size of the smallest boundary region $B$ such that the logical operator cannot be reconstructed from the complement of the boundary region $B$, which we denote by $B^{c}$. This means that the  distance $d_{x}$ of a logical algebra of a point $x$ is the volume of the smallest boundary region $B$ such that $x$ is not in the the entanglement wedge of $B^{c}$. That is,
\begin{align}
    d_{x} \, = \, \min_{B: x \notin \mathcal{E}(B^{c})} |B| \, , \label{eq-holo-d-point}
\end{align}
where $|B|$ is the volume of the region $B$. For a logical subalgebra associated to a finite bulk region $X$, the distance is given by 
\begin{align}
    d_{X} \, = \, \min_{x\in X} \, d_{x} \, . \label{eq-holo-d}
\end{align}

The price $p_{X}$ of a logical subalgebra corresponding to bulk region $X$ is the volume of the smallest boundary region on which any operator $\phi \in \mathcal{A}(X)$ can be represented. The subregion-subregion duality implies that the price is the volume of the smallest boundary region such that the region $X$ is in the entanglement wedge.  
\begin{align}
    p_{X} \, = \, \min_{B: X \in \mathcal{E}(B)} \, |B| \, . \label{eq-holo-p}
\end{align}
By comparing these definitions, one can deduce that $p_{X} \ge d_{X}$. This statement is called \textit{no free lunch} in \cite{Pastawski:2016qrs}.  

It is worthwhile to mention that the inequality of the no free lunch can be saturated for a bulk point $x$ if one assumes the notion of  \textit{geometric complementarity} \cite{Pastawski:2016qrs}. The geometric complementarity states that a bulk point is either in the entanglement wedge of a boundary region $B$ or in the entanglement wedge of the complementary region $B^{c}$. That is, if $x\notin \mathcal{E}(B^{c})$, then $x \in \mathcal{E}(B)$.  %Assuming the geometric complementarity, we deduce from Eq.~\eqref{eq-holo-d-point} and Eq.~\eqref{eq-holo-p} that the distance and price for a logical algebra of a single bulk point are the same. However, for general bulk regions, we have $p_{X} \ge d_{X}$.  

A stronger version of the no free lunch is the holographic strong Singleton bound. It states that the difference of price $p_{X}$ and distance $d_{X}$ of the logical algebra associated to a bulk region $X$ can not be less than the number of logical degrees of freedom $k_{X}$ in that bulk region \cite{Pastawski:2016qrs}. That is, 
\begin{align}
    k_{X} \, \le \, p_{X} - d_{X} \, .\label{eq-holo-singleton}
\end{align}

%In the following subsection, we consider a simple example of a holographic RG flow in which we find that there exists a finite bulk region for which the price and the distance of the corresponding logical algebra are the same. The holographic Singleton bound then implies that there should not be any logical degrees of freedom in that subregion. We then discuss in Sec.~(\ref{eq-sec-rw}) how to modify the definitions of the distance and the price to resolve this problem.  

%in Eq.~\eqref{eq-holo-d} and Eq.~\eqref{eq-holo-p} respectively so that the holographic Singleton bound does not become trivial.  %(ADD SOME WORDS ABOUT THE FOLLOWING SUBSECTION.)

\subsection{Holographic RG flow}

Consider the vacuum AdS$_{3}$ with the metric
\begin{align}
    ds^{2} \, = \, e^{2 r/L} \, \left( -dt^{2} + dx^{2} \right) \, + \, dr^{2} \, ,
\end{align}
where $L$ is the AdS length scale, and $r$ is the bulk radial coordinate. The boundary is located at $r = \infty$ and $r = -\infty$ is the Poincare horizon. This geometry is dual to the vacuum state of a $(1+1)$-dimensional CFT on a flat spacetime. The AdS length scale and the central charge of the CFT are famously related according to \cite{Brown:1986nw}
\begin{align}
    c \, = \, 3L/2G_{N} \, .\label{eq-cl-rel}
\end{align}

Now suppose we set off an RG flow on the CFT by deforming with a relevant operator. The geometry that is dual to the RG flow on the boundary is given by \cite{Girardello:1998pd,Freedman:1999gp,Girardello:1999bd}
\begin{align}
    ds^{2} \, = \, e^{2 A(r)} \, \left( -dt^{2} + dx^{2} \right) \, + \, dr^{2} \, ,
\end{align}
where $A(r)$ is such that $A(r) \, \sim \, {r/L_{UV}}$ near $r = \infty$ whereas $A(r) \, \sim \, {r/L_{IR}}$ near $r = - \infty$. $L_{UV}$ is related to the central charge of the UV theory according to Eq.~\eqref{eq-cl-rel}, whereas $L_{IR}$ is related to the central charge of the IR theory to which the UV theory flows \footnote{The holographic $c$-theorems say that the null energy condition in the bulk implies $L_{IR} < L_{UV}$ \cite{Myers:2010xs,Myers:2010tj}.}. The function $A(r)$ captures the flow of the boundary theory from UV to IR \cite{Myers:2010xs,Myers:2010tj}. 

In this work, we consider a simple example where $A(r)$ is given by a
\begin{align}
    A(r) \, = \, \begin{cases}
                                   r/L_{UV} & \quad r \geq 0 \\
      r/L_{IR} & \quad   r \leq 0 
    \end{cases} \, .
\end{align}
This simple model of holographic RG flows has been studied in \cite{Myers:2012ed,Albash:2011nq} where the minimal area surfaces corresponding to boundary regions are studied. Here, we review the results from \cite{Myers:2012ed}.

Consider a single interval on a boundary of size $\ell$. For small enough regions, the bulk stationary area surfaces remain near the boundary and do not penetrate to $r<0$ region (i.e. the IR region.) In particular, this type of stationary area surfaces only exist for $\ell \, \le \, \ell_{2} \, \equiv \,  2 L_{UV} \, $. For regions of length $\ell \, \ge \, \ell_{2} \, $, the stationary area surfaces must penetrate to the IR region. 

However, it was observed in \cite{Myers:2012ed} that even for $\ell \, < \, \ell_{2} \, $, there can exist stationary area surfaces that go to $r < 0$ region. In fact, such surfaces exist for $\ell \geq \ell_{1} \, $, where
\begin{align}
    \ell_{1} \, = \, 2L_{UV} \, \sqrt{1 - \left(1 - \frac{L_{IR}}{L_{UV}}\right)^{2} } \, .
\end{align}

Based on the above discussion, the HRT surfaces for $\ell \, < \, \ell_{1}$ completely stay in the UV region whereas the HRT surfaces for $\ell \, > \, \ell_{2}$ penetrate to the IR region. For intermediate size regions, both types of stationary area surfaces exist but the HRT surface is the one with a smaller area. It was observed in \cite{Myers:2012ed} that there exists a critical size of the interval, $\ell_{t}$, below which the surfaces that stay in the UV region has a smaller area and above which the surfaces that reaches the IR region has a smaller area. In the next section, we numerically calculate the critical size of the interval by comparing the area of the surfaces involved. The result of this analysis is shown as a plot of $\ell_{t}$ versus $L_{IR}/L_{UV}$ in (Fig.~\ref{fig-holo-rg} Right). 

Due to the `phase transition' in the HRT surface at $\ell = \ell_{t}$, there is a jump in the bulk entanglement wedge as well. When the size of the boundary interval is just smaller than $\ell_{t}$, the minimum radial point that is reached by the HRT surface is $r_{UV} > 0$. On the other hand, when the size of the boundary interval is just greater than $\ell_{t}$, the minimum radial point reached by the HRT surface is $r_{IR} < 0$. The jump in the entanglement wedge at $\ell = \ell_{t}$ can be measured in terms of the proper distance between these minimum radial points. We numerically calculate this proper distance %in the Appendix (\ref{appendix-holography})
and find that we can make this proper distance bigger by making the difference between $L_{IR}$ and $L_{UV}$ bigger; see (Fig.~\ref{fig-holo-rg} Right).

%\begin{figure}
%    \centering
%    \includegraphics[width=0.3\textwidth]{jump3.png}
%    \caption{\small{Pictorial representation of the phase transition in the HRT surfaces at $\ell = \ell_{t}$ for $L_{IR} = 0.3$ and $L_{UV} = 1.0$. The HRT surface for $\ell$ just bigger/smaller than $\ell_{t}$ is shown in blue/red color. The yellow shaded region is an example of a finite size region $X$ for which the distance and price are equal.}}
%    \label{fig-jump}
%\end{figure}

\begin{figure}
    \centering
    \includegraphics[width=0.4\textwidth]{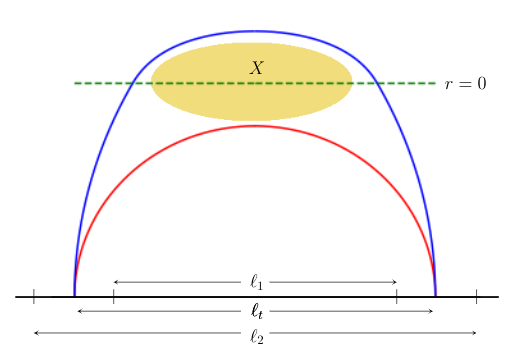}
    \caption{\small{Pictorial representation of the phase transition in the HRT surfaces at $\ell = \ell_{t}$ for $L_{IR} = 0.3$ and $L_{UV} = 1.0$. The critical length scale is in the range $\ell_{1} < \ell_{t} < \ell_{2}$. The HRT surface for $\ell$ just bigger/smaller than $\ell_{t}$ is shown in blue/red color. The yellow shaded region is an example of a finite size region $X$ for which the distance and price are equal.}}
    \label{fig-jump}
\end{figure}

The transition in the entanglement wedge at $\ell = \ell_{t}$ has interesting implications for holographic QECC. Consider a bulk region $X$ which is the intersection of the region $r < r_{UV} $ and the entanglement wedge of a boundary interval of size slightly greater than $\ell_{t}$. The condition $r < r_{UV} $ implies that the region $X$ is not in the entanglement wedge of any boundary interval of size less than $\ell_{t}$. Hence, according to Eqs.~\eqref{eq-holo-d-point}-\eqref{eq-holo-d}, the distance of the logical algebra associated to region $X$ is given by $\ell_{t}$. Moreover, according to Eq.~\eqref{eq-holo-p}, the price of the region $X$ is also given by $\ell_{t}$\footnote{The distance and the price are actually equal to $(\ell_{t})^{\alpha}$ where $\alpha = \log(2)/\log(\sqrt{2}+1)$ \cite{Pastawski:2016qrs}. This is the size of the fractal like disconnected intervals such that the entanglement wedge of the disconnected region has the same minimum radial point as the entanglement wedge of a single interval of size $\ell_{t}$. This construnction is called \textit{uberholography} in \cite{Pastawski:2016qrs}.}. This means that we have found a logical subalgebra associated to a finite size bulk region for which the price and the distance are the same. 
Comparing this with the holographic strong Singleton bound in Eq.~\eqref{eq-holo-singleton}, we deduce that the number of logical degrees of freedoms in that finite size subregion should be zero or else we get a violation of the holographic strong Singleton bound.  

We discuss in the next subsection how to modify the definition of the distance and the price to resolve this apparent paradox.

\subsection{calculation}

In this section, we present the details of the phase transition in the entanglement wedge that we discussed in Sec.~(\ref{sec:holography}). 

Consider a single interval on a boundary of size $\ell$. One possible stationary area surface anchored on this interval is the one that does not penetrate to $r<0$ region (i.e. the IR regrion). This surface is given by \footnote{This equation is only for the half of the surface. The center of the interval is chosen to be at $x=0$ and the surface is symmetric around $x=0$.}
\begin{align}
    x \, = \, L_{UV} \, \sqrt{ e^{-2r_{m}/L_{UV}} - e^{-2r/L_{UV}} } \, , \label{eq-surf-uv} %\right) \, ,
\end{align}
where $r_{m}$ is the minimum radial point that is reached by the minimal surface and is related to the size of the interval according to $\ell \, = \, 2L_{UV} e^{-r_{m}/L_{UV}} \, $. Note that this type of stationary area surface for which $r_{m}>0$ can only exist for $\ell \, \leq \, \ell_{2} \, $, where $\ell_{2} \, = \, 2L_{UV}$.
%Therefore, the maximum size of the interval for which this type of minimal surface can exist is $\ell_{2} \, = \, 2L_{UV} \, $, which can be determined taking $r_{m} = 0$.

For intervals of size $\ell > \ell_{2}$, the stationary area surfaces discussed above would penetrate to the IR region. However, it was observed in \cite{Myers:2012ed} that even for $\ell \, \leq \, \ell_{2} \, $, there can exist stationary area surfaces that go to $r < 0$ region. In fact, there can be two such surfaces for a given $\ell$ and they can be written as $x = x_{\pm}(r)$ where
\begin{eqnarray}
    &&x_{\pm}(r) \, = \nn\\
    &&\, \begin{cases}
                                   L_{IR} \sqrt{K_{\pm}^{2} -  e^{-2r/L_{IR}} },\:  &r_{m,\pm} \leq r \leq 0 \nn\\
      L_{UV} \left( \sqrt{K_{\pm}^{2} - e^{-2r/L_{UV}} } - K_{\pm}\right) + \frac{\ell}{2}  &    r \geq 0 \ .
    \end{cases} \, . \label{eq-surf-ir}
\end{eqnarray}
The minimum radial point $r_{m,\pm}$ is related to $K_{\pm}$ according to $K_{\pm} \, = \, e^{-r_{m,\pm}/L_{IR}} $. The continuity of $x_{\pm}(r)$ at $r = 0$ implies
\begin{align}
    \ell \, = \, 2L_{UV} K_{\pm} \, - \, 2\left(L_{UV}-L_{IR}\right) \, \sqrt{K_{\pm}^{2} \, - \, 1 } \, ,
\end{align}
which can be inverted to get
\begin{align}
    K_{\pm} =  \frac{ L_{UV} \ell  \pm  \left(L_{UV}-L_{IR}\right)  \sqrt{ \ell^{2} + 4L_{IR} \left( L_{IR} -2 L_{UV}\right) } }{2L_{UV}^{2} - 2\left(L_{UV}-L_{IR}\right)^{2} }  . \label{eq-kpm}
\end{align}
These surfaces can only exist when $K_{\pm}$ are real-valued which requires $\ell \geq \ell_{1} \, $, where
\begin{align}
    \ell_{1} \, = \, 2L_{UV} \, \sqrt{1 - \left(1 - \frac{L_{IR}}{L_{UV}}\right)^{2} } \, .
\end{align}

\begin{figure}
    \centering
    \includegraphics[width=0.23\textwidth]{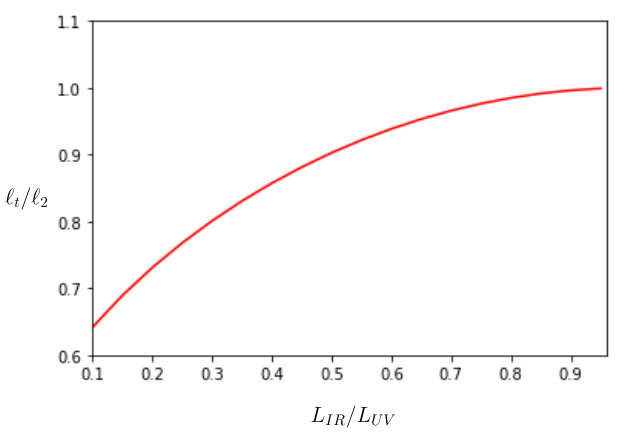}
    \includegraphics[width=0.23\textwidth]{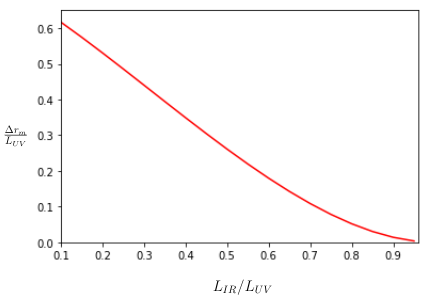}
    \caption{\small{Numerical calculation of the (Left) critical length, $\ell_{t}$, and (Right) the jump in the entanglement wedge at the phase transition, $\Delta r_{m}$, as a function of $L_{IR}/L_{UV}$.}}
    \label{fig-holo-rg}
\end{figure}

Therefore, there are three possible stationary area surfaces corresponding to a boundary interval of size $\ell_{1} \leq \ell \leq \ell_{2}$. The HRT surface, however, is the one with the smallest area among these three surfaces. Let us denote the area of the surface in Eq.~\eqref{eq-surf-uv} by $A_{0}$ and the area of surfaces $x_{\pm}(r)$ in Eq.~\eqref{eq-surf-ir} by $A_{\pm}$. Then $A_{0}$ is given by \cite{Ryu:2006bv}%[CITE RT]
\begin{align}
    A_{0} \, = \, 2L_{UV} \, \log \left(\frac{\ell}{\delta_{UV}}\right) \, ,
\end{align}
whereas $A_{\pm}$ are given by \cite{Myers:2012ed}
\begin{align}
    A_{\pm} \, =& \, \, 2L_{UV} \, \log\left( \frac{2L_{UV}K_{\pm}}{\delta_{UV}} \right) \, \nonumber\\ -& \, \, \left(L_{UV}-L_{IR}\right) \, \log\left[ \frac{ K_{\pm} \, + \, \sqrt{K_{\pm}^{2}-1} }{ K_{\pm} \, - \, \sqrt{K_{\pm}^{2}-1} } \right] \, ,
\end{align}
where $\delta_{UV}$ is a UV cutoff of the boundary theory. It was observed in \cite{Myers:2012ed} that the area $A_{-}$ is always greater than $A_{+}$ and that there exists a critical size of the interval, $\ell_{t}$, below which $A_{0}<A_{+}$ and above which $A_{+}<A_{0} $\footnote{Even though the areas $A_{1}$, $A_{+}$, and $A_{-}$ depend on the UV cutoff, the difference of any two areas is independent of the cutoff. This makes the comparison of the areas meaningful.}. This means that for some critical $l_t$, when $\ell < \ell_{t}$ the surface in Eq.~\eqref{eq-surf-uv} is the RT surface, whereas when $\ell > \ell_{t}$ the surface $x_{+}$ in Eq.~\eqref{eq-surf-ir} is the RT surface. We can calculate the critical size of the interval by numerically solving $A_{0}-A_{+} = 0$. The result of this analysis is shown as a plot of $\ell_{t}$ versus $L_{IR}/L_{UV}$ in (Fig.~\ref{fig-holo-rg} Right). As we can see from the figure, $\ell_{t}$ increases with $L_{IR}/L_{UV}$ and it approaches $\ell_{2}$ when $L_{IR}$ approaches $L_{UV}$. 

% \begin{figure}
%     \centering
%      \begin{subfigure}[b]{0.4\textwidth}
%          \centering
%          \includegraphics[width=\textwidth]{2lt.png}
%           \caption{ }
%           \label{fig-holo-lt}
%       \end{subfigure}
%       \quad \hfill
%       \begin{subfigure}[b]{\textwidth}
%           \centering
%          \includegraphics[width=.4s\textwidth]{2delta_rm.png}
%          \caption{ }
%          \label{fig-holo-rm}
%     \end{subfigure}
%     \caption{\small{Numerical calculation of the (Left) critical length, $\ell_{t}$, and (Right) the jump in the entanglement wedge at the phase transition, $\Delta r_{m}$, as a function of $L_{IR}/L_{UV}$.}}
%     \label{fig-holo-rg}
% \end{figure}

As we discussed in Sec.~(\ref{sec:holography}), there is a jump in the entanglement wedge due to the phase transition in the HRT surfaces.  The jump in the entanglement wedge at $\ell = \ell_{t}$ can be measured in terms of the proper distance between the minimum points reached by the two candidate HRT surfaces. This proper distance is given by
\begin{align}
    \Delta r_{m} \, = \, r_{UV} \, - \, r_{IR} \, .
\end{align}
where $r_{UV} \, = \, -L_{UV} \, \log\left(\frac{\ell_{t}}{2L_{UV}}\right) \, $, and $r_{IR} \, = \, - L_{IR} \, \log K_{+}(\ell_{t})$, and $K_{+}(\ell_{t})$ can be determined using Eq.~\eqref{eq-kpm}. 
We plot $\Delta r_{m}$ as a function of $L_{IR}/L_{UV}$ in (Fig.~\ref{fig-holo-rg} Right). As we can see from the figure, the jump in the HRT surfaces, and hence in entanglement wedges, is more significant when the difference between $L_{IR}$ and $L_{UV}$ is large.% A pictorial representation of the jump in the HRT surfaces is shown in Fig.~(\ref{fig-jump}).  

\subsection{Price, distance, and the reconstruction wedge} \label{sec:eq-sec-rw}

%solution: redefine the bound with RW. how?

We observed in the previous subsection that a phase transition in the entanglement wedge led us to a violation of the holographic strong Singleton bound. In this subsection, we discuss this violation can be resolved. 

The definition of the entanglement wedge in terms of the minimal area surface is only valid at the leading order in $O(1/G_{N})$. At subleading order in $G_{N}$, we have to take the entanglement entropy of the bulk quantum fields into consideration \cite{Faulkner:2013ana,Engelhardt:2014gca}. More precisely, the entanglement wedge of a boundary region $B$ is the domain of dependence of a bulk region between $B$ and a codimension-$2$ surface of stationary generalized entropy. The generalized entropy of a region is equal to the area of the boundary of that region (in Planck's units) plus the entropy of the quantum fields in the region. This subleading correction, as is recently emphasized \cite{Hayden:2018khn,Akers:2019wxj,Akers:2020pmf}, can be significant when there is a phase transition in the minimal surfaces. 

Now suppose there is a state $\rho_{X}$ of the quantum fields in region $X$ which we introduced in the previous subsection. If this state is pure, then the discussion of the entanglement wedge is unchanged and we end up with a phase transition at $\ell = \ell_{t}$. However, when the state $\rho_{X}$ is mixed, then there is no phase transition and the entanglement wedge at $\ell = \ell_{t}$ is region between the boundary interval and the red surface shown in (Fig.~\ref{fig-jump}). Hence, the entanglement wedge depend on the state $\rho_{X}$. 

The dependence of state on entanglement wedge has been recently discussed in \cite{Hayden:2018khn,Akers:2019wxj,Akers:2020pmf,wang2021refined}. In particular, it was argued in \cite{Akers:2019wxj} that the bulk region that can be reconstructed given a boundary region $B$ is not the entanglement wedge of $B$. In fact, this region can be macroscopically smaller than the entanglement wedge, $\mathcal{E}(B)$. The \textit{reconstruction wedge}, $\mathcal{R}(B)$, corresponding to the boundary region $B$ is defined to be the intersection of all the entanglement wedges of $B$ for every state in the code space \cite{Akers:2019wxj}.

In the example that we discussed above, the reconstruction wedge of boundary interval of size $\ell=\ell_{t}$ is the entanglement wedge when the state $\rho_{X}$ is mixed. Hence, there is no phase transition in the reconstruction wedge at $\ell = \ell_{t}$.

With the fact that the reconstruction wedge is smaller than the entanglement wedge, we restate one of the main statements in the main text. %Let us start by discussing price. Since the price of a logical subalgebra is a smallest boundary region on which any logical operator can be represented, it is natural to define price in terms of the reconstruction wedge rather than the entanglement wedge. More precisely, 
To solve the violation of holographic Singleton bound, we propose that the definition of the price for a logical algebra associated to a bulk region $X$ should be modified from Eq.~\eqref{eq-holo-p} to 
\begin{align}
    p_{X} \, = \, \min_{B: X \in \mathcal{R}(B)} \, |B| \, . \label{eq-holo-p-R}
\end{align}
Similarly, %the distance of a logical subalgebra is the smallest boundary region $B$ such that the logical algebra cannot be reconstructed from the complement boundary region, $B^{c}$. Again, it is more natural to define the distance in terms of the reconstruction wedge. Therefore, 
we propose that the distance of a subregion $X$ is
\begin{align}
    d_{X} \, = \, \min_{x \in X} \, d_{x} \, ;  \quad\quad\quad\,  d_{x} \, = \, \min_{B: x \notin \mathcal{R}(B^{c})} |B| \, . \label{eq-holo-d-R}
\end{align}

With the new definition, although the distance is still determined by the $\ell = \ell_{t}$, the price is determined by $\ell = \ell_{2}$ which is the largest length for which the surface in Eq.~\eqref{eq-surf-uv} exists\footnote{Again, the distance and the price are given by $(\ell_{t})^{\alpha}$ and $(\ell_{2})^{\alpha}$ respectively where $\alpha = \log(2)/\log(\sqrt{2}+1)$ as determined by the uberholography construnction.}. 

%These modifications to the definition of the price and the distance resolve the violation to the holographic Singleton bound that we discussed in the previous subsection. For the region $X$ that we defined, the distance is still determined by the $\ell = \ell_{t}$. The price, on the other hand, is determined by $\ell = \ell_{2}$ which is the largest length for which the surface in Eq.~\eqref{eq-surf-uv} exists\footnote{Again, the distance and the price are given by $(\ell_{t})^{\alpha}$ and $(\ell_{2})^{\alpha}$ respectively where $\alpha = \log(2)/\log(\sqrt{2}+1)$ as determined by the uberholography construnction.}. 

%This finishes our discussion of the holographic quantum error correction. 

\bibliography{main}